\documentclass[10pt,epsfig]{article} 
\pdfoutput=1
\usepackage{pstricks}
\usepackage{enumerate}
\usepackage{float}
\usepackage{subfig}
\usepackage{color}
\usepackage{slashed}
\usepackage[makeroom]{cancel}

\usepackage{amssymb, amsmath,mathrsfs}
\usepackage{graphicx}
\usepackage{multirow}
\usepackage{cite}
\usepackage{chngcntr}
\usepackage[colorlinks=true,
            linkcolor=red,
            urlcolor=blue,
            citecolor=blue]{hyperref}
\usepackage{dcolumn}
\usepackage{bm}
\usepackage{color}
 \usepackage{epstopdf}
\long\def\rpl#1!!#2!!{\textcolor{red}{#1} \textcolor{blue}{#2}}
\def\baselinestretch{1.3}
\newcommand{\ba}{\begin{array}}
\newcommand{\ea}{\end{array}}
\newcommand{\bd}{\begin{displaymath}}
\newcommand{\ed}{\end{displaymath}}
\newcommand{\besub}{\begin{subequations}}
\newcommand{\eesub}{\end{subequations}}
\newcommand{\be}{\begin{equation}}
\newcommand{\ee}{\end{equation}}
\newcommand{\bea}{\begin{eqnarray}}
\newcommand{\eea}{\end{eqnarray}}
\newcommand{\no}{\nonumber\\}

\def\a{\alpha}

\def\b{\beta}
\def\g{\gamma}

\def\l{\lambda}

\def\L{\Lambda}

\def\q2 {q^2}

\def\bt{\begin{table}}
\def\et{\end{table}}

\catcode`@=11 % This allows us to modify PLAIN macros.
\def \gsim{\mathrel{\mathpalette\@versim>}}
\def \lsim{\mathrel{\mathpalette\@versim<}}
\def \@versim#1#2{\lower0.4ex\vbox{\baselineskip\z@skip\lineskip\z@skip
     \lineskiplimit\z@\ialign{$\m@th#1\hfil##\hfil$
     \crcr#2\crcr\sim\crcr}}}
\catcode`@=12 % at signs are no longer letters

\allowdisplaybreaks
\begin{document}

%THE TEXT STARTS HERE
\begin{flushright}
{HRI-RECAPP-2017-003}
\end{flushright}

\begin{center}

{\large \textbf {High-scale validity of a two Higgs doublet scenario: predicting collider signals}}\\[15mm]

Nabarun Chakrabarty$^{\dagger}$\footnote{nabarunc@hri.res.in} and Biswarup Mukhopadhyaya$^{\star}$\footnote{biswarup@hri.res.in}  \\
{\em Regional Centre for Accelerator-based Particle Physics \\
     Harish-Chandra Research Institute, HBNI,\\
 Chhatnag Road, Jhunsi, Allahabad - 211 019, India}\\[5mm] 

\end{center}

\begin{abstract} 

It is possible to ameliorate the Higgs vacuum stability problem by switching over to two Higgs doublet models (2HDM), ensuring a stable electroweak vacuum  up to the Planck scale, even though
the top quark mass may be on the high side. However, the simultaneous requirements of perturbative unitarity, and also compatibility with collider and flavour data, constrain the parameter space severely. We investigate the collider signals answering to the regions allowed by such constraints. In particular, the near degeneracy of the neutral heavy scalar and the pseudoscalar is a feature that is probed. The LHC allows distinguishability of these two states, together with signal significance of at least 3$\sigma$, in its high-luminosity run. While $e^+ e^-$ colliders may have rather low event rates, muon colliders, cashing on the principle of radiative return, can probe 2HDM scenarios with (pseudo)scalar masses up to a TeV or so, though with the price of losing distinction between the CP-even and odd states.

\end{abstract}

\newpage
\setcounter{footnote}{0}

\def\baselinestretch{1.5}
\counterwithin{equation}{section}
%==========================================================================
%==========================================================================

\section{Introduction}\label{Intro}

The spin-0 particle of mass around 125 GeV discovered by the ATLAS~\cite{Aad:2012tfa} and CMS~\cite{Chatrchyan:2012xdj} collaborations at the Large Hadron Collider (LHC)
apparently completes the particle spectrum of the Standard Model (SM).
Moreover, the couplings of this particle to the other SM particles are progressively getting closer
to the corresponding SM values. However, issues ranging from the presence of dark matter in the universe to the naturalness problem of the electroweak scale keep alive the hope of finding physics beyond the SM (BSM). While the search for such new physics remains on, a rather pertinent question  is to ask is whether the SM by itself can ensure vacuum stability at scales above that of electroweak symmetry
breaking (EWSB). This is because the Higgs quartic coupling evolving via SM interactions alone tends to turn negative
in between the Electroweak (EW) and Planck scales, thereby making the scalar potential unbounded from
below. This is particularly true if the top quark mass is on the upper edge of its allowed band~\cite{Degrassi:2012ry,Buttazzo:2013uya}. While a \emph{metastable} EW minimum remains a possibility, stabilising the EW vacuum calls for the introduction of additional bosonic fields
preferably by extending the SM Higgs sector. A number of scenarios comprising new physics are suggested for retrieving vacuum stability, a representative list of which is~\cite{PhysRevD.86.043511,Gonderinger2010,Chen:2012faa,PhysRevD.78.085005,Chakrabortty2016361,
He:2013tla,PhysRevD.92.055002}. One important demonstration in this context is that stability till the Planck scale is 
restored, irrespective of the top-mass uncertainty, just by switching over to two Higgs doublet models (2HDM). 2HDMs open up a world of enriched collider phenomenology, CP-violation from the scalar sector and also dark matter candidates in special cases.

However, a challenge faced while 
alleviating the vacuum instability problem using 2HDMs (or any extended Higgs sector for that matter) is that the quartic couplings so introduced tend to become non-perturbative while
evolving under renormalisation group (RG). A balance between these two extremes is struck through judicious boundary conditions, which in turn leads to strong constraints on the masses and mixing
angles. Elaborate accounts of this can be found in recent works. Two important points emerge from such studies. First, the spectrum of the non-standard scalars
allows for only a small splitting. Secondly, the couplings of the 125 GeV Higgs with gauge bosons
should have rather small deviation from the SM values. On the other hand, the gauge interactions of the non-standard scalars become suppressed. 

In this work, we aim to investigate the observability of a 2HDM at 
the present\cite{PhysRevD.90.015008,Keus2016,Kanemura2014524,Basso:2012st,PhysRevD.94.095005,
Patrick:2016rtw,Li:2016umm,Arhrib:2016rlj,Akeroyd:2016ymd} and upcoming colliders\cite{LopezVal:2009qy,PhysRevD.88.115003} within the parameter region that allows for high-scale 
validity (including both vacuum stability and perturbativity). This could turn challenging since the search prospects could be severely inhibited
by the constraints. For instance, to discern a 2HDM from the SM background through
resonances, fully reconstructible final states need to be looked at. The corresponding event rates tend to be small, owing to the constraints on the interaction strengths
that come from demanding the dual requirement of high scale vaccum stability and perturbative unitarity. Moreover, removal of the backgrounds requires event selection criteria which further lower the signal strength.

To be more specific, the CP-even heavier neutral Higgs could lead to a four-lepton
cascade at the LHC via the $ZZ$ state. Side by side, the CP-odd scalar leaves its signature in the completely reconstructible channel $hZ$ where $h$ denotes the SM-like Higgs. The two final states mentioned above are indicative of the opposite CP-properties of the decaying Higgses, which from our requirement, are destined to have closely spaced masses. We adopt a cut-based analysis to calculate the statistical significance in the respective signals. We perform this analysis for both Type-I and Type-II 2HDM. The allowed parameter space for the latter scenario is obtained via extensive investigation in reference~\cite{Chakrabarty:2014aya}. For the former, though an analysis is found in~\cite{Das:2015mwa}, for the sake of completeness, we present a set of results here that go beyond what has been reported. It is found that the constraints from flavour changing neutral current (FCNC)
phenomena put a strong lower limit on the Type-II 2HDM charged scalar mass (and, via the correlation demanded by high-scale validity, on the heavy neutral scalar and pseudoscalar masses as well). Thus while obtaining LHC signals, the region of the parameter space in the Type-II case is relatively more restricted. Keeping this in mind, we also present a brief discussion on the prospects at other types of colliders. In particular, we find that muon colliders can be useful in this respect.

This study comprises of the following parts. In section~\ref{model}, we briefly review
the 2HDM and survey its candidature as a UV-complete scenario. Section~\ref{param} highlights
the intrinsic features of the parameter space that permits high-scale stability. The search prospects at the LHC, and, future leptonic colliders are elaborated in sections~\ref{lhc} and
\ref{other} respectively. We summarise our findings and conclude in section~\ref{con}.

\section{2HDM and high scale validity.}\label{model}

Type-I and II 2HDMs, as well as the constraints on them have already been discussed in literature \cite{Branco:2011iw}. We present a small resume here for completeness.
We consider the most general renormalizable
scalar potential for two doublets $\Phi_1$ and $\Phi_2$, each having
hypercharge $(+1)$:
\bea
V(\Phi_1,\Phi_2) &=&
m^2_{11}\, \Phi_1^\dagger \Phi_1
+ m^2_{22}\, \Phi_2^\dagger \Phi_2 -
 m^2_{12}\, \left(\Phi_1^\dagger \Phi_2 + \Phi_2^\dagger \Phi_1\right)
+ \frac{\lambda_1}{2} \left( \Phi_1^\dagger \Phi_1 \right)^2
+ \frac{\lambda_2}{2} \left( \Phi_2^\dagger \Phi_2 \right)^2
\no & &
+ \lambda_3\, \Phi_1^\dagger \Phi_1\, \Phi_2^\dagger \Phi_2
+ \lambda_4\, \Phi_1^\dagger \Phi_2\, \Phi_2^\dagger \Phi_1
+ \frac{\lambda_5}{2} \left[
\left( \Phi_1^\dagger\Phi_2 \right)^2
+ \left( \Phi_2^\dagger\Phi_1 \right)^2 \right]
\no & &
+\lambda_6\, \Phi_1^\dagger \Phi_1\, \left(\Phi_1^\dagger\Phi_2 + \Phi_2^\dagger\Phi_1\right) + \lambda_7\, \Phi_2^\dagger \Phi_2\, \left(\Phi_1^\dagger\Phi_2 + \Phi_2^\dagger\Phi_1\right).
\label{treepot}
\eea

We parametrise the doublets as
\be
\Phi_{i} = \frac{1}{\sqrt{2}} \begin{pmatrix}
\sqrt{2} w_i^{+} \\
v_i + h_i + i z_i
\end{pmatrix}~ \rm{for}~\textit{i} = 1, 2.
\label{e:doublet}
\ee
One defines tan$\beta = \frac{v_2}{v_1}$.
 In such a case, the scalar spectrum consists of a pair of neutral CP even scalars ($h,H$), a CP odd neutral scalar ($A$) and a charged scalar ($H^+$). The mass matrices are brought into diagonal form
by the action unitary matrices comprising of mixing angles $\a$ and $\b$. This scenario in general allows for CP-violation in the scalar sector~\cite{Grzadkowski:2013rza,Shu:2013uua,PhysRevD.72.095002}, through the phases in  $m_{12}$ and $\lambda_5$. However, this would lead to a contamination of our proposed search channels due to interfernece 
efects coming from $H-A$ mixing. Thus we restrict ourselves to a CP conserving scenario only.

A particular fermion generation can couple to both $\Phi_1$ and $\Phi_2$ in a 2HDM without 
violating the gauge symmetry. However, this leads to flavour changing neutral 
currents (FCNC) mediated by the Higgses, which are tightly constrained by experimental data.
A manner to annul the FCNCs is to adhere to specific schemes of Yukawa interations\cite{PhysRevD.15.1966,PhysRevD.15.1958}, that
are consequences of discrete symmetries. A example is the $\mathbb{Z}_2$ 
symmetry under which $\Phi_1 \rightarrow -\Phi_1$ and $\Phi_2 \rightarrow \Phi_2$. This 
demands $m_{12},\l_6,\l_7 = 0$. Assigning
appropriate $\mathbb{Z}_2$ charges to the fermions
 gives rise to the celebrated Type-I and Type-II models~\cite{Branco:2011iw}. 
While the primary motivation of the above is to suppress flavour changing neutral 
currents (FCNC)~\cite{Kim2015,Crivellin:2013wna,Baum:2008qm}, it reduces the number of free parameters in the Yukawa sector.\footnote{It was reported in
~\cite{PhysRevD.58.116003} that the FCNCs are stable under Renormalisation Group.} This also simplifies
the expressions for the one-loop beta functions. Note that one could introduce $Z_2$ violation 
in the scalar potential only. This would ultimately lead to FCNC, however which would 
be radiatively suppressed. In this study, we consider both the cases of an exactly $\mathbb{Z}_2$ symmetric 2HDM, and one that violates it in the scalar potential.

We choose $\{\text{tan}\beta,m_h,m_H,m_A,m_{H^+},m_{12},c_{\beta -\alpha},\l_6,\l_7\}$ as the
set of independent input parameters. The rest of the quartic couplings are expressed in terms of which for 
convenience. With $v = 246 ~\rm GeV$ and writing $c_{\alpha}$ = cos$\alpha$, $s_{\alpha}$ = sin$\alpha$, the remaining couplings can be expressed as
\besub
\bea
\label{e:l1}
\l_1 &=& \frac{1}{v^2 c^2_\b}~\Big(c^2_\a m^2_H + v^2 s^2_\a m^2_h - m^2_{12}\frac{s_\b}{c_\b} - \frac{3}{2} \l_6 v^2 s_{\b} c_{\b} - \frac{1}{2} \l_7 v^2 \frac{s^3_{\b}}{c_{\b}} \Big),\\
\label{e:l2}
\l_2 &=& \frac{1}{v^2 s^2_\b}~\Big(s^2_\a m^2_H + v^2 c^2_\a m^2_h - m^2_{12}\frac{c_\b}{s_\b} - \frac{3}{2} \l_7 v^2 s_{\b} c_{\b} - \frac{1}{2} \l_6 v^2 \frac{c^3_{\b}}{s_{\b}}\Big),\\
\label{e:l4}
\l_4 &=& \frac{1}{v^2}~(m^2_A - 2 m^2_{H^+}) + \frac{m^2_{12}}{v^2 s_\b c_\b}
 - \frac{1}{2 t_{\b}} \l_6 - \frac{1}{2} t_{\b} \l_7,\\
\label{e:l5}
\l_5 &=& \frac{m^2_{12}}{v^2 s_\b c_\b} - \frac{m^2_A}{v^2}
 - \frac{1}{2 t_{\b}} \l_6 - \frac{1}{2} t_{\b} \l_7, \\
\label{e:l3}
\l_3 &=& \frac{1}{v^2 s_\b c_\b}((m^2_H - m^2_h)s_\a c_\a + m^2_A s_\b c_\b - \l_6 v^2 c^2_\b - \l_7 v^2 s^2_\b) - \l_4.
\eea
\label{e:Couplings}
\eesub
The mass parameters $m_{11}$ and $m_{22}$ in the scalar potential are traded off using the EWSB conditions. 
A given set of input parameters serves as boundary conditions for $\l_i$ 
for the analysis using RG equations. While carrying out the analysis, several constraints coming from both theory and
experiments must be satisfied.

\subsubsection{Perturbativity, unitarity and vacuum stability}
For the 2HDM to remain a perturbative theory at a given
energy scale, one requires $\lvert \lambda_{i} \rvert
\leq 4\pi~(i=1,\ldots,5)$ and $ \lvert y_{i} \rvert \leq
\sqrt{4\pi}~(i=t,b,\tau)$ at that scale. This translates
into upper bounds on the model parameters at low as well as 
high energy scales.

%A more sophisticated version of such bounds comes from the the requirement of partial wave unitarity in longitudinal gauge boson scattering.
The matrix containing 2$\rightarrow$2 scattering amplitudes of longitudinal gauge bosons
can be mapped to a corresponding matrix for the scattering of the goldstone 
bosons\cite{Akeroyd:2000wc,Horejsi:2005da,Kanemura:2015ska,Ginzburg:2005dt}, by virtue of the EW equivalence theorem. The theory is deemed unitary if each eigenvalue of the aforementioned amplitude matrix does not exceed 8$\pi$. The expressions for the eigenvalues are given below.
\besub
\bea
a_{\pm}&=& \frac32(\lambda_1+\lambda_2)\pm \sqrt{\frac94 (\lambda_1-\lambda_2)^2+(2\lambda_3+\lambda_4)^2},\\
b_{\pm}&=& \frac12(\lambda_1+\lambda_2)\pm \sqrt{\frac14 (\lambda_1-\lambda_2)^2+\lambda_4^2},\\
c_{\pm}&=& d_{\pm} = \frac12(\lambda_1+\lambda_2)\pm \sqrt{\frac14 (\lambda_1-\lambda_2)^2+\lambda_5^2},\\
e_1&=&(\lambda_3 +2\lambda_4 -3\lambda_5),\\
e_2&=&(\lambda_3 -\lambda_5),\\
f_1&=& f_2 = (\lambda_3 +\lambda_4),\\
f_{+}&=& (\lambda_3 +2\lambda_4 +3\lambda_5),\\
f_{-}&=& (\lambda_3 +\lambda_5).
\eea
\label{e:LQTeval}
\eesub
When the quartic part of the scalar
potential preserves CP and $\mathbb{Z}_2$ symmetries, the aforementioned
eigenvalues are discussed in
\cite{Kanemura:1993hm,Akeroyd:2000wc,Horejsi:2005da}.

Demanding high-scale positivity of the 2HDM potential along various directions in the field space
leads to the following conditions on the scalar potential
\cite{Branco:2011iw,Ferreira:2004yd,PhysRevD.18.2574,Nie199989}:
\besub 
\bea
\label{e:vsc1}
\rm{vsc1}&:&~~~\lambda_{1} > 0, \\
\label{e:vsc2}
\rm{vsc2}&:&~~~\lambda_{2} > 0, \\
\label{e:vsc3}
\rm{vsc3}&:&~~~\lambda_{3} + \sqrt{\lambda_{1} \lambda_{2}} > 0, \\
\label{e:vsc4}
\rm{vsc4}&:&~~~\lambda_{3} + \lambda_{4} - |\lambda_{5}| + \sqrt{\lambda_{1} \lambda_{2}} > 0.
\label{e:vsc5}
\eea
\label{eq:vsc}
\eesub
Meeting the above positivity criteria at each scale of evolution 
effectively rules out deeper vacua at high energy scales.

In addition to the above, the spliting amongst the scalar masses is restricted by invoking the $T$-parameter constraint. We have used $\Delta T = 0.05 \pm 0.12$ following \cite{Baak:2014ora}, where $\Delta T$ measures departure from
the SM contribution. We have filtered all points in our parameter space through the above constraints and retained only those points that negotiate it successfully. Measurement of the rate for $b \rightarrow s \gamma$ leads to $m_{H^+} \geq 480$ GeV in case 
of the Type-II 2HDM\cite{Mahmoudi:2009zx,Olive:2016xmw}. In case of Type-I, there is no such lower bound. The constraint $m_{H^+} \geq 80$ GeV originating from direct searches however still persists.

\section{Type-I 2HDM: Allowed parameter space for stable vacuum}\label{param}

We start by completing the existing studies~\cite{Eberhardt:2014kaa,Chakrabarty:2014aya,Das:2015mwa,Ferreira:2015rha} on the parameter space allowing for high scale vaccum stability and perturbativity for a Type I 2HDM. A corresponding discussion for the Type-II 2HDM can be seen in \cite{Chakrabarty:2014aya}.
We fix $m_h = 125$ GeV and $M_t = 175$ GeV, the rest of the parameters are generated randomly  in the following ranges.\\ 
 $\rm tan\beta \in [1,20]$, $m_H \in [200,1000]$, $m_A \in [200,1000]$, $m_{H^+} \in [200,1000]$, $\rm cos(\beta - \alpha) \in [-0.4,0.4]$, $\l_6 \in [-1,1]$, $\l_7 \in [-1,1]$.
The generated values of the masses and mixing angles are translated to the basis of the quartic
couplings using Eqs.~\ref{e:l1} to ~\ref{e:l3}.

\begin{figure} %[t]
\begin{center}
\includegraphics[scale=0.40]{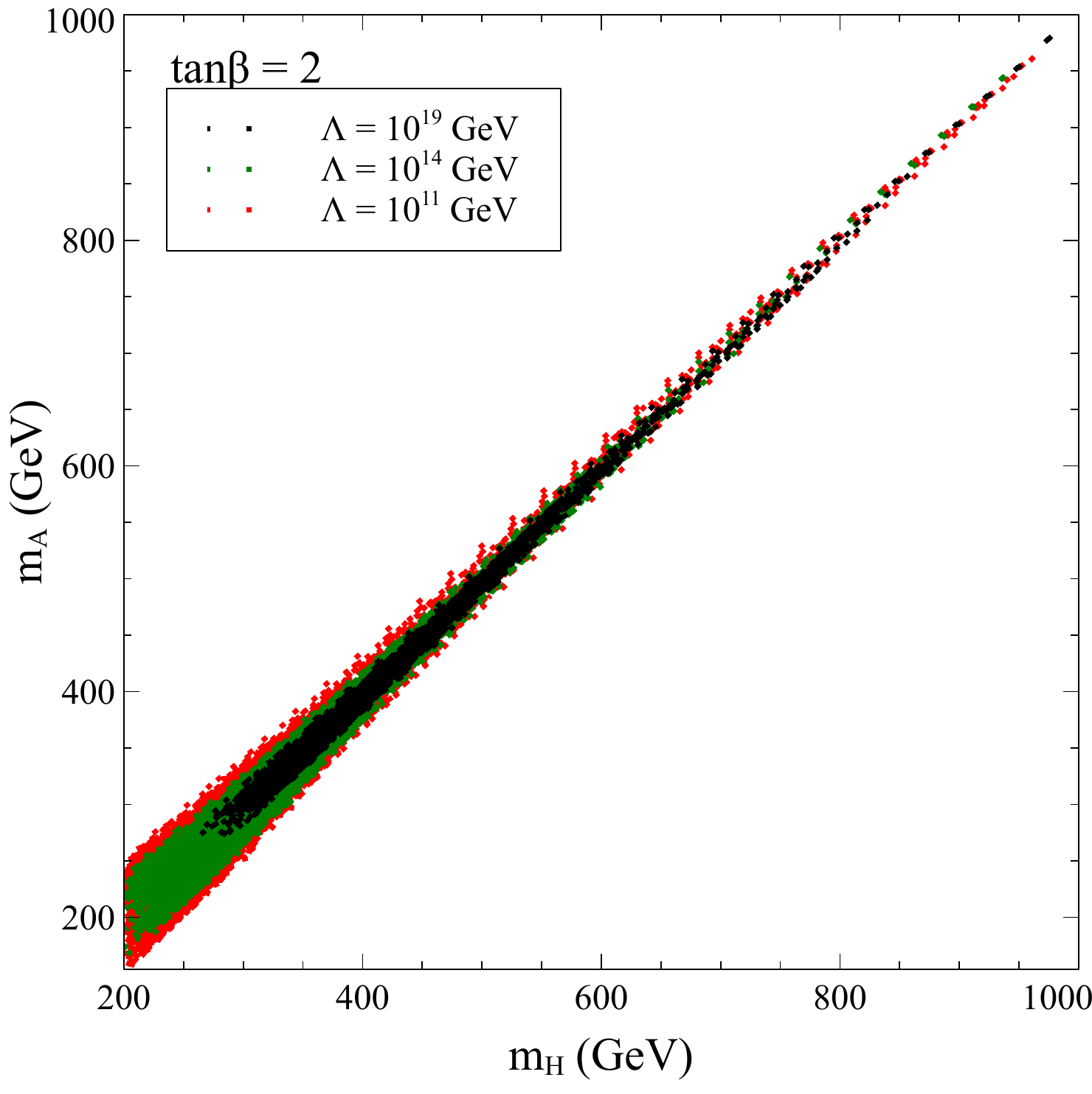}~~~ 
\includegraphics[scale=0.40]{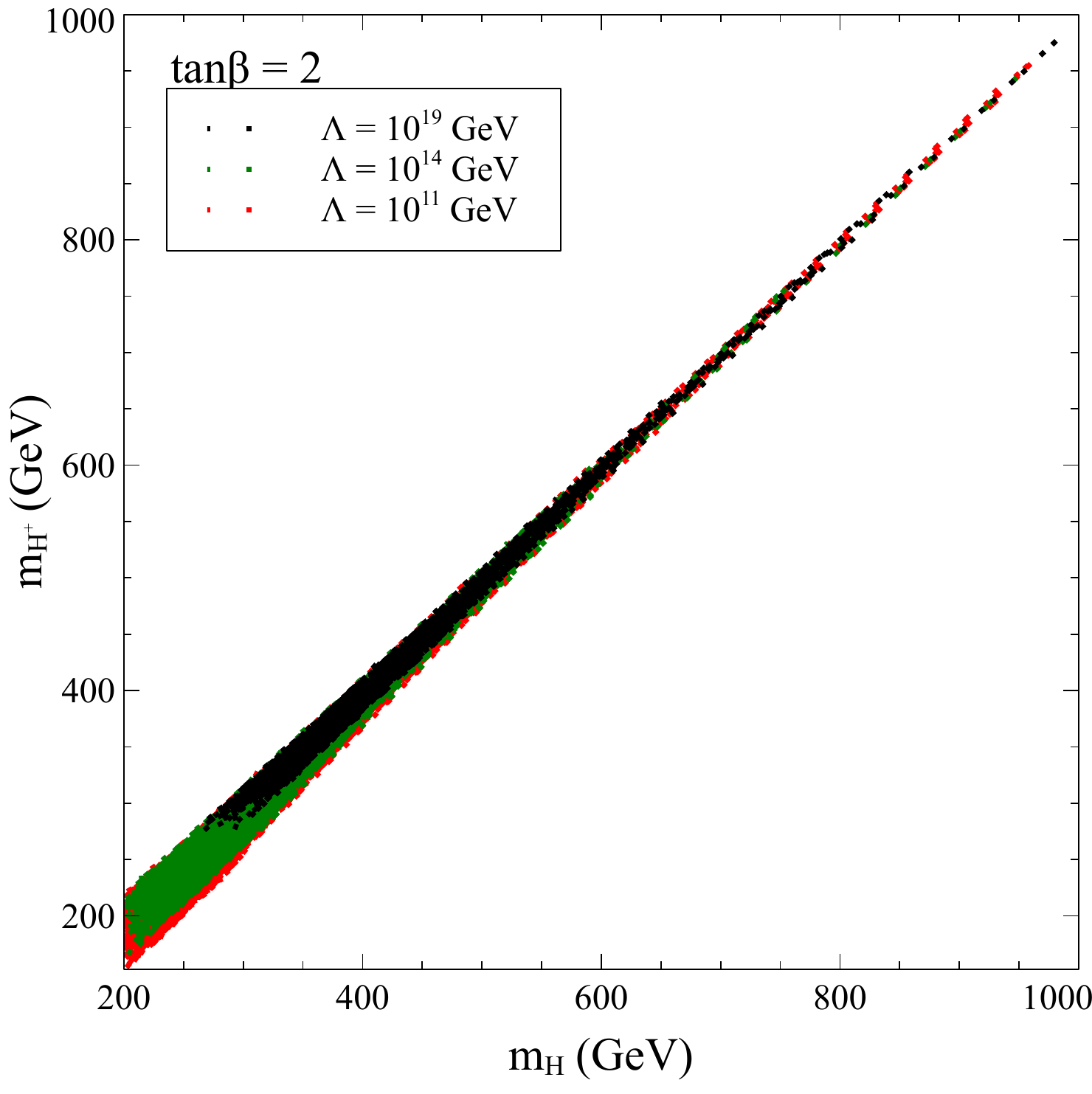} 
\includegraphics[scale=0.40]{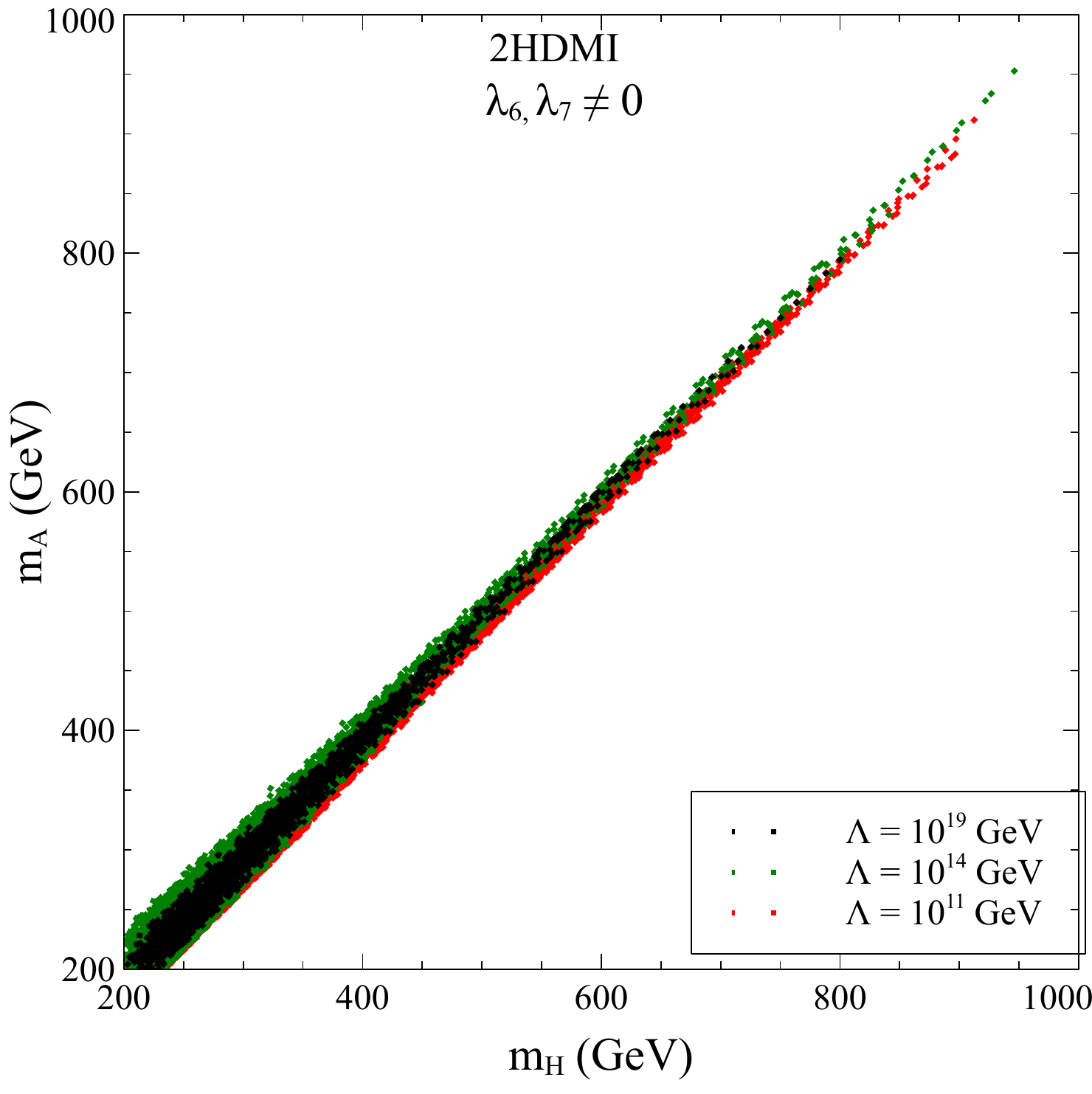}~~~ 
\includegraphics[scale=0.40]{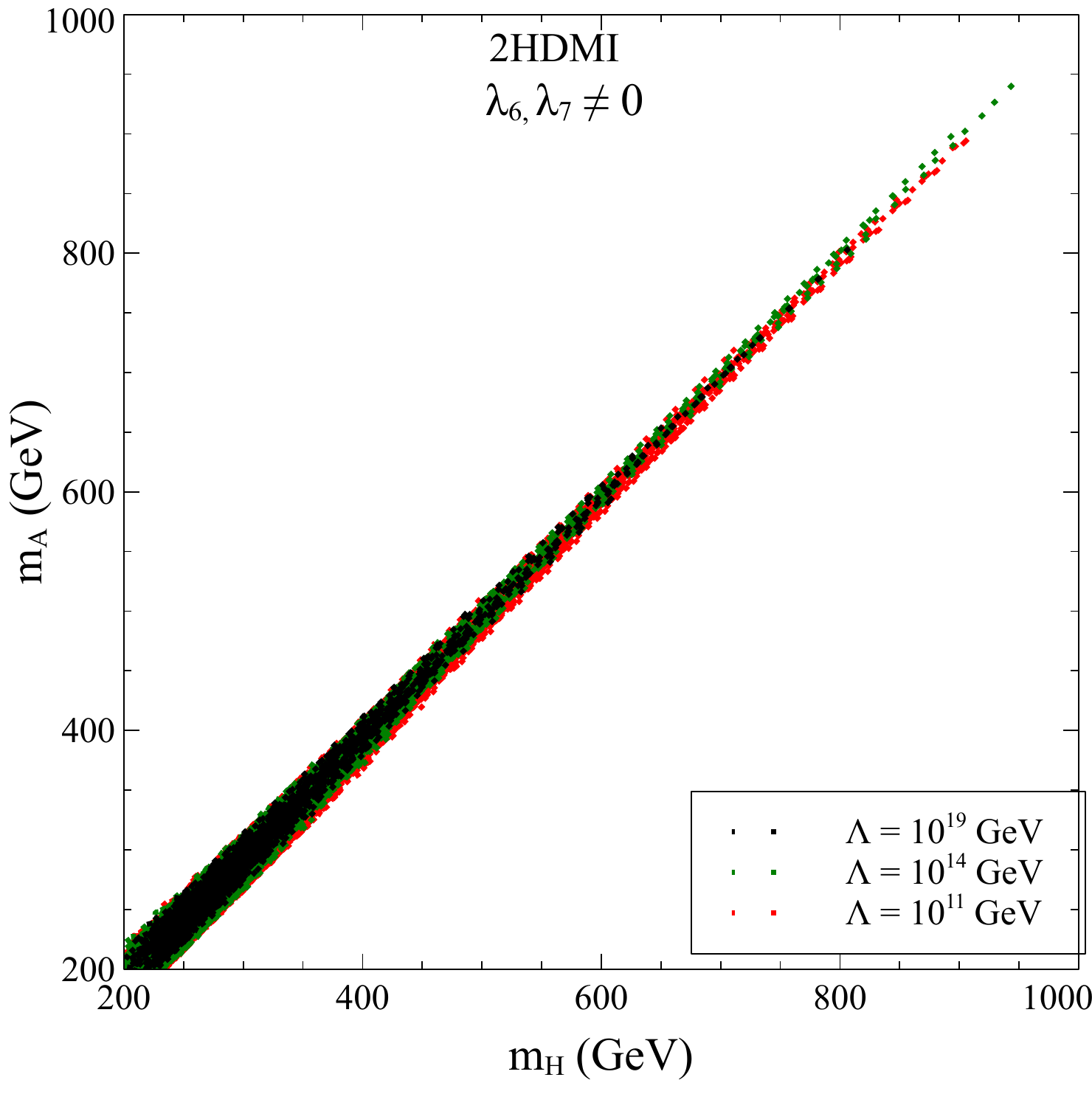} 
\caption{Distribution of the parameter points valid till $\Lambda$ in the $m_H-m_A$ (left) and $m_H-m_{H^+}$ (right) planes for the Type-I 2HDM. The colour coding can be read from the legends. We fix tan$\beta$ = 2 as a benchmark. The upper(lower) plots correspond to $\l_6 = \l_7 = 0$($\l_6 ,\l_7 \neq 0$). We have varied $\l_6,\l_7$ in the interval [-1,1] for the lower plots.}
\label{f:m-m}
\end{center}
\end{figure}

The strong 
correlation among the masses, namely $m_H \simeq m_A \simeq m_{H^+}$, is revealed from Fig.~\ref{f:m-m}. This itself can be traced back to Eqs.~\ref{e:l1} to ~\ref{e:l3}. Any large mass gap results in giving large values for $\l_i$ at the EWSB scale itself, such that they turn non-perturbative rather early in the course of evolution.
This feature is also corroborated in\cite{Das:2015mwa}. It is important to note that the mass-splitting depends, albeit weakly, on the chosen
 value of tan$\beta$. For instance, in case of tan$\beta = 2$, the maximum splitting allowed is
$\simeq 15$ GeV for $\L = 10^{19}$ GeV. This goes down to $\simeq 10$ GeV in case of tan$\beta = 10$ for the same value of $\L$. It should be noted here that the bound on mass splitting that comes from the requirement of perturbativity till high scales is much more stringent than what is obtained by the imposition of the T-parameter constraint alone.

Also important is the ensuing constraint on cos($\beta - \alpha$) which decides
the interaction strengths between $W,Z$ and the non-standard scalars. The more suppressed is
cos($\b - \a$), closer are the $h$-interactions to the corresponding values. Thus, measurement
of signal strengths of $h$ leads to constraint on this parameter\cite{Bernon:2015qea,Cheon:2012rh,PhysRevD.90.095006}.
Models valid up to $10^{19}$ GeV
could allow for $|\rm cos(\beta - \alpha)|$ $\leq$ 0.15 and $|\rm cos(\beta - \alpha)|$ $\leq$ 0.05
for tan$\beta$ = 2 and tan$\beta$ = 10 respectively. This bound can be amply relaxed by choosing
a lower $\L$, for example one finds $|\rm cos(\beta - \alpha)|$ $\leq$ 0.14 in case of tan$\beta$ = 10 if one demands validity up to $10^{19}$ GeV. This apparent correlation between the UV cut-off scale and the maximum allowed value of $\rm cos(\beta - \alpha)$, could lead us to predict the maximal extrapolation scale up to which such a 2HDM could be probed at the colliders. Of course, such a correlation can be noticed for the Type-II scenario as well. The additional result presented here, over and above what is found in the literature, is the establishment of the mass correlations for $\l_6,\l_7 \neq$ 0, as shown in Fig.1.

\section{Signals at the LHC: Types I and II.}\label{lhc}  
The previous section illustrates that higher the UV cutoff of a 2HDM is, 
tighter become the mass-splitting and the bound on $|\rm cos(\beta - \alpha)|$.
Such a constrained scenario makes its observability at the LHC a rather challenging task,
as also emphasized in section~\ref{Intro}.
In particular, if we probe $H$ and $A$ via their decays into recontructible final states,
then the invariant mass distributions of the decay products would coincide. However,
probing $H$ and $A$ in reconstructible but distinct final states could enable one to
tag the CP of the decaying boson. Given that, we propose the following signals:\\
(i) $p p \longrightarrow H \longrightarrow Z Z \longrightarrow 4 l$ \\
(ii) $p p \longrightarrow A \longrightarrow h Z \longrightarrow l^+ l^- b b$

We have implemented the model using \texttt{FeynRules}\cite{Alloul20142250}. The generated
Universal FeynRules Output (UFO) files are then fed to the Monte-Carlo (MC)
event generator MadGraph ~\cite{Alwall:2014hca} for generation of event samples. The parton-showering
and hadronisation is carried out in the \texttt{PYTHIA-6} \cite{1126-6708-2006-05-026} framework. We
simulated $H \rm ~and ~A$ production through the gluon-gluon fusion (ggF) channel using the CTEQ6L1 parton distribution functions. 
This is because ggF offers higher rates compared to other channels. The renormalisation and factorisation scales have been set at $m_H$ and $m_A$ for the first and second signals respectively.
We mention in this
context that detector simulation and analysis of the events were done using \texttt{Delphes}\cite{deFavereau2014}.

For simulating the proposed final states, we hold $m_H$ and $m_A$ fixed and scan
over the remaining input quantities. From the randomly generated parameter sets,
we select an illustrative assortment of benchmark points (Table~\ref{Benchmark}) to highlight
the main findings of the analysis. 

\begin{table}[h]

\centering
\begin{tabular}{|c c c c c|}
\hline
Benchmark  & $m_{H}$(GeV) & $m_{A}$(GeV) & $m_{12}$(GeV)  & cos($\b - \a$) \\ \hline \hline
BP1a  & 350 & 351 & 200 & -0.18 \\
BP1b  & 350 & 351 & 200 & -0.12 \\ \hline
BP2a  & 400 & 401 & 230 & -0.15 \\
BP2b  & 400 & 401 & 230 & -0.10 \\ \hline
BP3a  & 500 & 501 & 280 & -0.095 \\
BP3b & 500 & 501 & 280 & -0.070 \\
BP3c & 500 & 501 & 280 & -0.050 \\ \hline
BP4a  & 550 & 551 & 320 & -0.075 \\
BP4b  & 550 & 551 & 320 & -0.060 \\
BP4c  & 550 & 551 & 320 & -0.050 \\ \hline
BP5a  & 600 & 601 & 350 & -0.050 \\
BP5b  & 600 & 601 & 350 & -0.035 \\
BP5c  & 600 & 601 & 350 & -0.025 \\ \hline
\end{tabular}
\caption{Benchmarks chosen for simulating the proposed channels. We have taken $m_h = 125$ GeV and tan$\beta$ = 2.5 throughout. Any higher tan$\beta$ would to a lower ggF rate and so was not chosen.}
\label{Benchmark}
\end{table}

The benchmarks are distict from another \emph{vis-a-vis} RG evolution patterns. While choosing them, it was ensured that the UV-cutoff of a given benchmark does not change upon switching between the Type-I and Type-II models. For instance, in the case where $\l_6 = \l_7$ = 0, BP1b, BP1b, BP3c, BP4c and BP5c are conservative input sets ensuring a stable
vaccum and a perturbative model till $\sim 10^{19}$ GeV. This can be read from the small values
of $|\rm cos(\beta - \alpha)|$ characterizing them. The other benchmarks are however not that conservative, but stil they manage to stabilise the vaccum till at least $10^{11}$ GeV. Likeweise,
BP3b and BP4b are included to estimate the statistical significance of scenarios valid till $10^{14}$ GeV. For a given set of couplings, elevating the masses of $H$ and $A$ progressively diminishes the intensity of the signals, and, also narrows the allowed band of $|\rm cos(\beta - \alpha)|$. The choice of the  benchmarks is thus guided by the aim to understand the maximum $m_H, m_A$ as well as the highest UV cut-off up to which the scenario can be experimentally observed.

\subsubsection{$p p \longrightarrow H \longrightarrow Z Z \longrightarrow 4 l$}

$H_2$ is produced through gluon fusion and decays to two on-shell $Z$ bosons. We look for a final state where the $Z$ bosons subsequently decay into four leptons\cite{Aad:2015kna}. The dominant background for this process
comes from $ZZ(^*)$ production. Taking into account subleading contributions from the $Z\g$ and $\g \g$ channels and multiplying by appropriate next-to-leading order (NLO) K-factors \cite{Alwall:2014hca}, the total background cross section is $\simeq$ 42 fb. Some basic cuts, as listed below, were applied during event generation.

\textbf{Basic-cuts:}
\begin{itemize}
\item All leptons have a minimum transverse momentum of 10 GeV, $p_T^{l} \geq 10$ GeV.
\item Pseudorapidity of the leptons must lie within the window$|\eta^{l}| \leq 2.5$.
\item All possible lepton-pairs are resolved using $\Delta R_{ll} > 0.3$.
\end{itemize}

We multiply the ggF cross sections of $H$ production by an NLO K factor of 1.5.
The cuts listed below were further imposed.

\textbf{Selection cuts:}
\begin{itemize}
\item \textbf{SC1}: The invariant mass of the final state leptons lie within the window $m_H - 15 \rm ~GeV \leq m_{4l} \leq m_H$ + \rm 15 GeV.
\item \textbf{SC2}: The transverse momenta of the leptons lie above the thresholds $p_T^{l_1} > p_{T,\rm min}^{l_1}$, $p_T^{l_2} > p_{T,\rm min}^{l_2}$, 
$p_T^{l_3} > 30$ \rm GeV, $p_T^{l_4} > 20$ \rm GeV.
\item \textbf{SC3}: Transverse momenta of the reconstructed $Z$-bosons satisfy $p_T^{Z_1} > p_{T, \rm min}^{Z_1}$, $p_T^{Z_2} > p_{T, \rm min}^{Z_2}$.
\end{itemize}

We take $p_T^{Z_1/Z_2} = 20, 20, 40, 50, 70$ GeV and $\{p_{T,\rm min}^{l_1}, p_{T,\rm min}^{l_2}\}$ =  $\{50 \rm ~GeV,30 \rm ~GeV\}$, $\{50 \rm ~GeV,30 \rm ~GeV\}$, $\{80 \rm ~GeV,50 \rm ~GeV\}$, $\{90 \rm ~GeV,70 \rm ~GeV\}$, $\{100 \rm ~GeV, 70 \rm ~GeV\}$ for BP1, BP2, BP3, BP4, BP5 respectively, the decisive factor in this choice of $p_T^{Z_1/Z_2}$ being $m_H$, for any benchmark point.

For $m_H > 500$ GeV,
the leading and the subleading leptons are strongly boosted, thus having a good probability of suviving the
strong $p_T$ cuts. In addition, appropriate cuts on the $p_T$ of the Z-bosons also contributes
towards improving the signal-to-background ratio. Denoting the number of signal and background events as $\mathcal{N}_S$ and $\mathcal{N}_B$ 
at a given integrated luminosity ($\mathcal{L}$), the statistical significance or the confidence
limit (CL) is defined as CL = $\frac{\mathcal{N}_S}{\sqrt{\mathcal{N}_S + \mathcal{N}_B}}$.

\begin{table}[h]
\centering
\begin{tabular}{|c c c c c c c c c|}
\hline
Benchmark   & $\sigma^{SC}_{S}$ (fb) & $\sigma^{SC}_{B}$ (fb) & $\mathcal{N}_S^{100}$ & $\mathcal{N}_B^{100}$ & $\mathcal{N}_S^{3000}$ & $\mathcal{N}_B^{3000}$ & $\rm CL_{100}$ & $\rm CL_{3000}$\\ \hline \hline
BP1a  & 0.173 & 0.334 & 17.36 & 33.40 & 520.94 & 1002.18 & 2.43 & 13.34\\
BP1b  & 0.145 & 0.334 & 14.54 & 33.40 & 436.31 & 1002.18 & 2.10 & 11.503 \\ \hline
BP2a  & 0.104 & 0.194 & 10.42 & 19.46 & 312.73 & 584.00 & 1.90 & 10.44 \\ 
BP2b  & 0.071 & 0.194 & 7.11 & 19.46 & 213.38 & 584.00 & 1.37 & 7.55 \\ \hline
BP3a  & 0.026 &  0.064 &  2.59 &  6.48 &  77.99 &  194.60 &  0.86 &  4.72\\
BP3b  & 0.016 &  0.064 &  1.68 &  6.48 &  50.52 &  194.60 &  0.58 &  3.22
\\
BP3c  &  0.009 &  0.064 &  0.97 &  6.48 &  29.37 &  194.60 &  0.35 &  1.96
\\ \hline
BP4a  & 0.011  & 0.041  & 1.13  & 4.16  & 34.06  & 124.91  & 0.49 &  2.70\\
BP4b & 0.008 &  0.041 &  0.81 &  4.16 &  24.52 &  124.91 &  0.36 &  2.00\\
BP4c  & 0.006 &  0.041 &  0.61 &  4.16 &  18.33 &  124.91 &  0.27 &  1.53\\ \hline
BP5a  & 0.004 & 0.029 & 0.41 & 2.96 & 12.32 & 89.090347 & 0.22 & 1.22\\
BP5b  & 0.002 & 0.029 & 0.22 & 2.96 & 6.70 & 89.090347 & 0.12 & 0.68\\
BP5c  & 0.001 & 0.029 & 0.12 & 2.96 & 3.61 & 89.090347 & 0.06 & 0.37\\ \hline
\end{tabular}
\caption{A record of the number of surviving events in the $H \rightarrow 4 l$ channel after the selection cuts at the $\sqrt{s} = 14 $ TeV LHC for a Type-I 2HDM. Here $\mathcal{N}_S^{100(3000)}$ and $\mathcal{N}_B^{100(3000)}$ and  respectively denote the number of events $\mathcal{L} = 100(3000)$ $\rm fb^{-1}$. Besides,
$\rm CL_{100(3000)}$ denotes the confindence level at $\mathcal{L} = 100(3000)$ $\rm fb^{-1}$.
}
\label{4l_TypeI}
\end{table}

\begin{table}[h]
\centering
\begin{tabular}{|c c c c c c c c c|}
\hline
Benchmark   & $\sigma^{SC}_{S}$ (fb) & $\sigma^{SC}_{B}$ (fb) & $\mathcal{N}_S^{100}$ & $\mathcal{N}_B^{100}$ & $\mathcal{N}_S^{3000}$ & $\mathcal{N}_B^{3000}$ & $\rm CL_{100}$ & $\rm CL_{3000}$\\ \hline \hline
BP3a  & 0.025 & 0.064 & 2.56 & 6.48 & 76.99 & 194.60 & 0.85 & 4.67 \\
BP3b  &  0.016 &  0.064 &  1.65 &  6.48 &  49.65 &  194.60 &  0.58 &  3.17
\\
BP3c  &  0.009 &  0.064 &  0.95 &  6.48 &  28.73 &  194.60 &  0.35 &  1.92\\ \hline
BP4a  & 0.011 &  0.041 &  1.12 &  4.16 &  33.64 &  124.91  & 0.48  & 2.67 \\
BP4b  & 0.008 &  0.041 &  0.80 &  4.16 &  24.15 &  124.91 &  0.36 &  1.97\\
BP4c & 0.006 &  0.041 &  0.60 &  4.16 &  18.02 &  124.91 &  0.27 &  1.50\\ \hline
BP5a  & 0.004 & 0.029 & 0.40 & 2.96 & 12.148 & 89.09 & 0.22 & 1.20\\
BP5b  & 0.002 & 0.029 & 0.21 & 2.96 & 6.58 & 89.09 & 0.124 & 0.67\\
BP5c & 0.001 & 0.029 & 0.11 & 2.96 & 3.54 & 89.09 & 0.06 & 0.36
\\ \hline
\end{tabular}
\caption{A record of the number of surviving events in the $H \rightarrow 4 l$ channel after the selection cuts at the $\sqrt{s} = 14 $ TeV LHC for a Type-II 2HDM. Here $\mathcal{N}_S^{100(3000)}$ and $\mathcal{N}_B^{100(3000)}$ and  respectively denote the number of events $\mathcal{L} = 100(3000)$ $\rm fb^{-1}$. Besides,
$\rm CL_{100(3000)}$ denotes the confindence level at $\mathcal{L} = 100(3000)$ $\rm fb^{-1}$.}
\label{4l_TypeII}
\end{table}

\begin{figure} %[t]
\begin{center}
\includegraphics[scale=0.48]{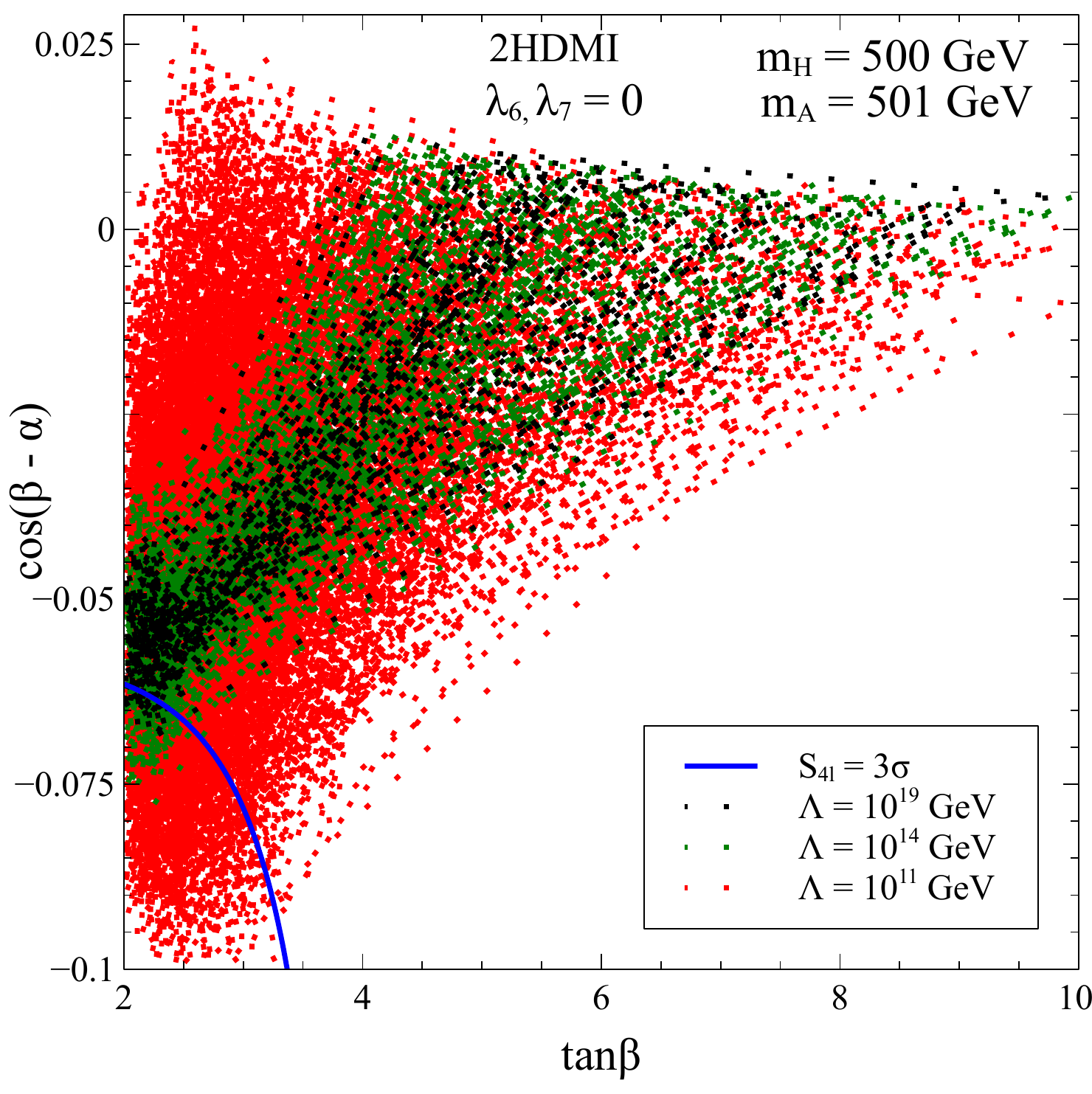}~~~ 
\includegraphics[scale=0.48]{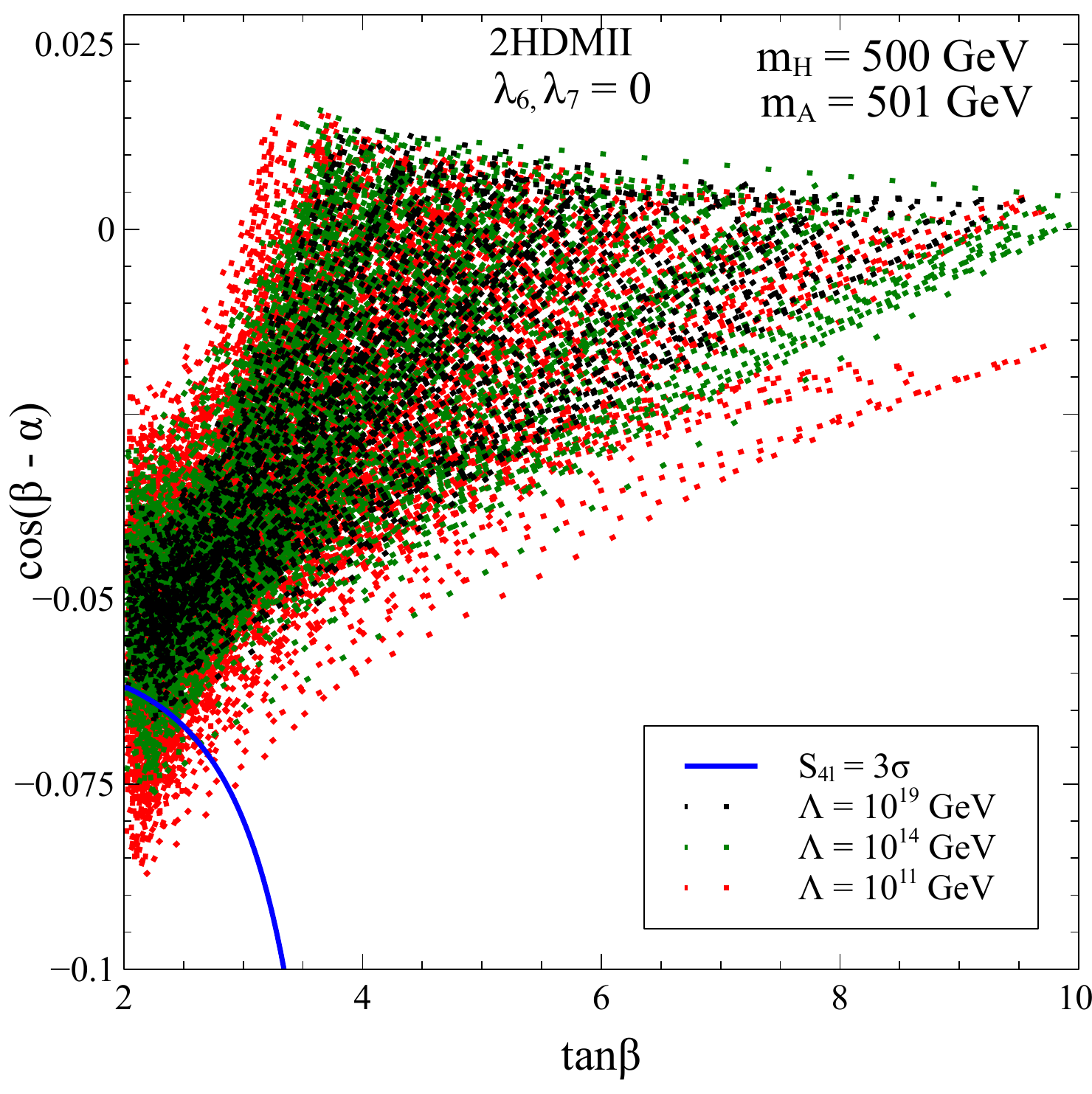}
\includegraphics[scale=0.48]{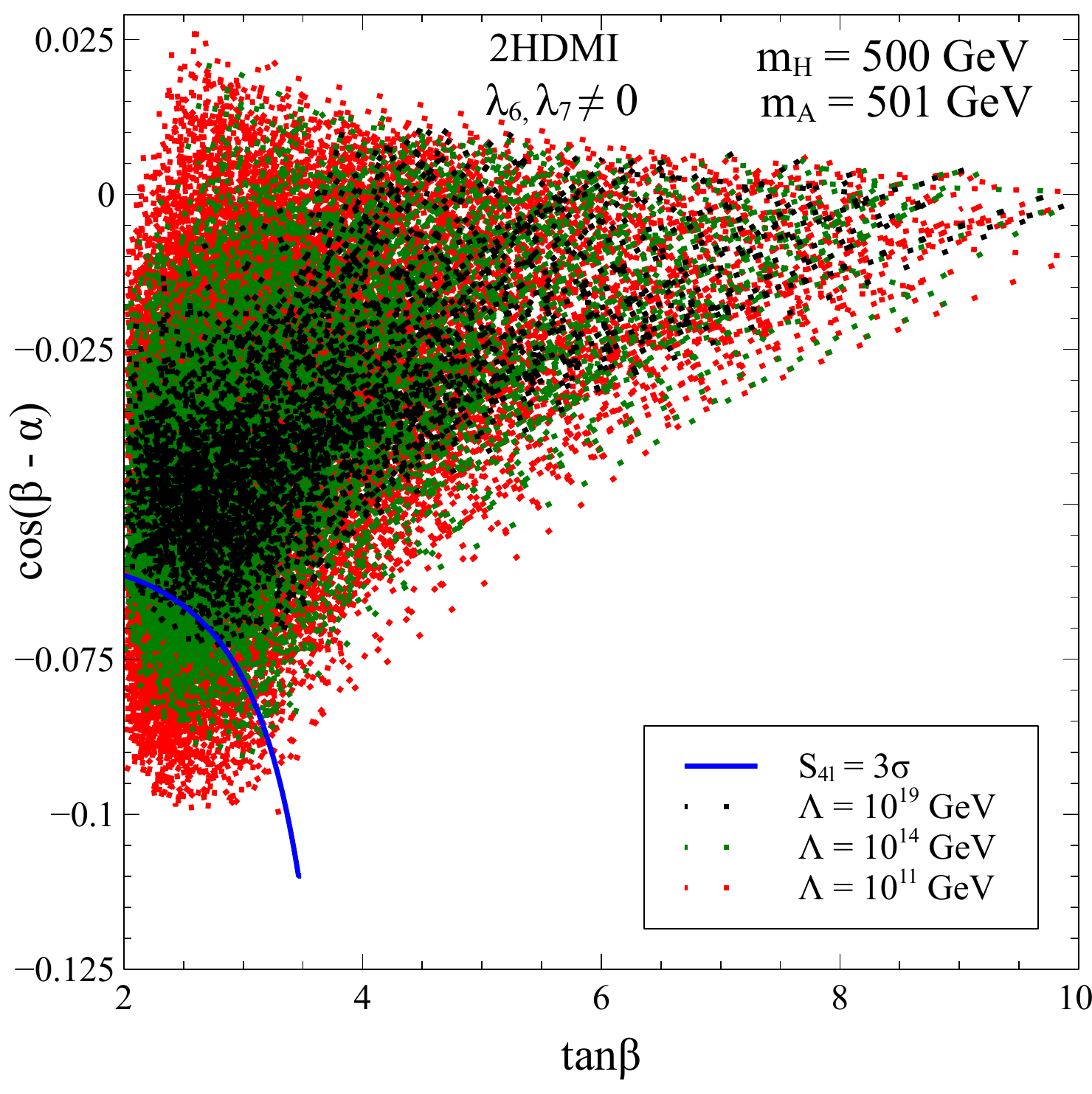}~~~ 
\includegraphics[scale=0.48]{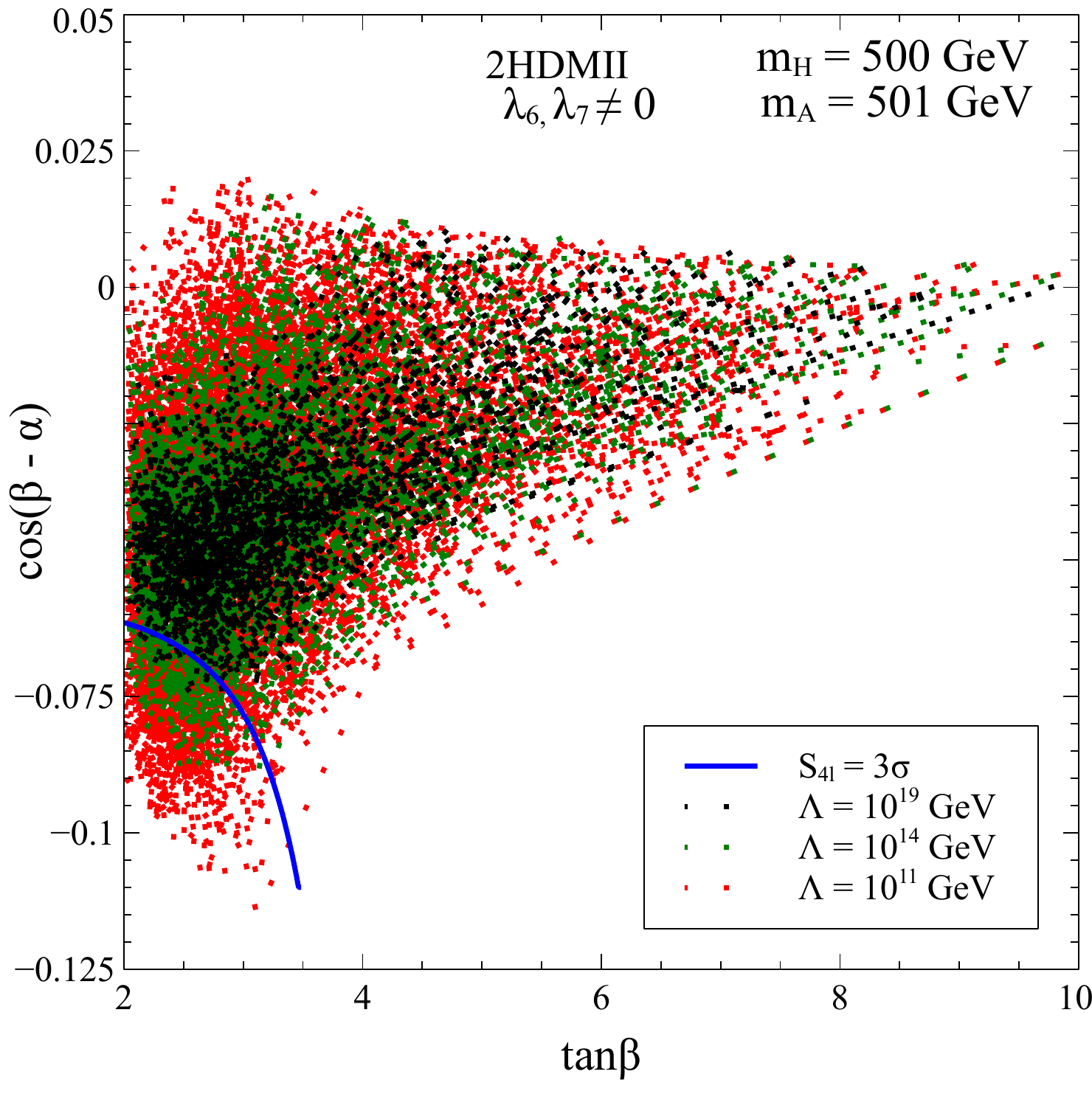}  
\caption{The paramater space in the tan$\beta$ vs. $c_{\b - \a}$ plane for $m_H = 500 \rm ~GeV$ and $m_A = 501$ GeV that allows for validity till $10^{11}$ GeV(red), $10^{14}$ GeV(green) and $10^{19}$ GeV(black). The region inside the blue curve corresponds to a signal significance greater than or equal to 3$\sigma$. 
The upper and lower plots are for $\l_6 = \l_7 = 0$ and $\l_6,\l_7 \neq 0$ respectively.}
\label{4l_500}
\end{center}
\end{figure}

\begin{figure} %[t]
\begin{center}
\includegraphics[scale=0.48]{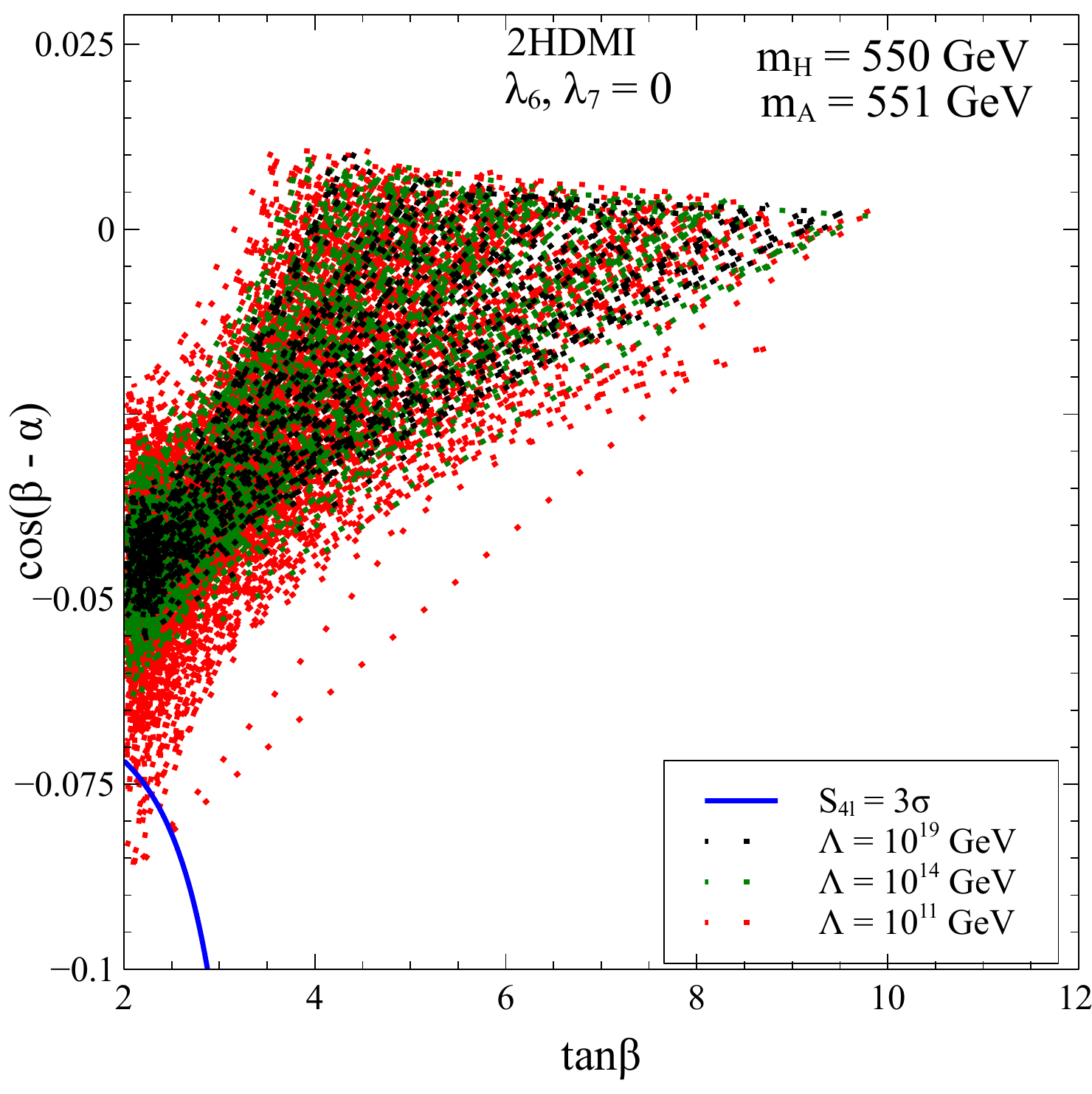}~~~
\includegraphics[scale=0.48]{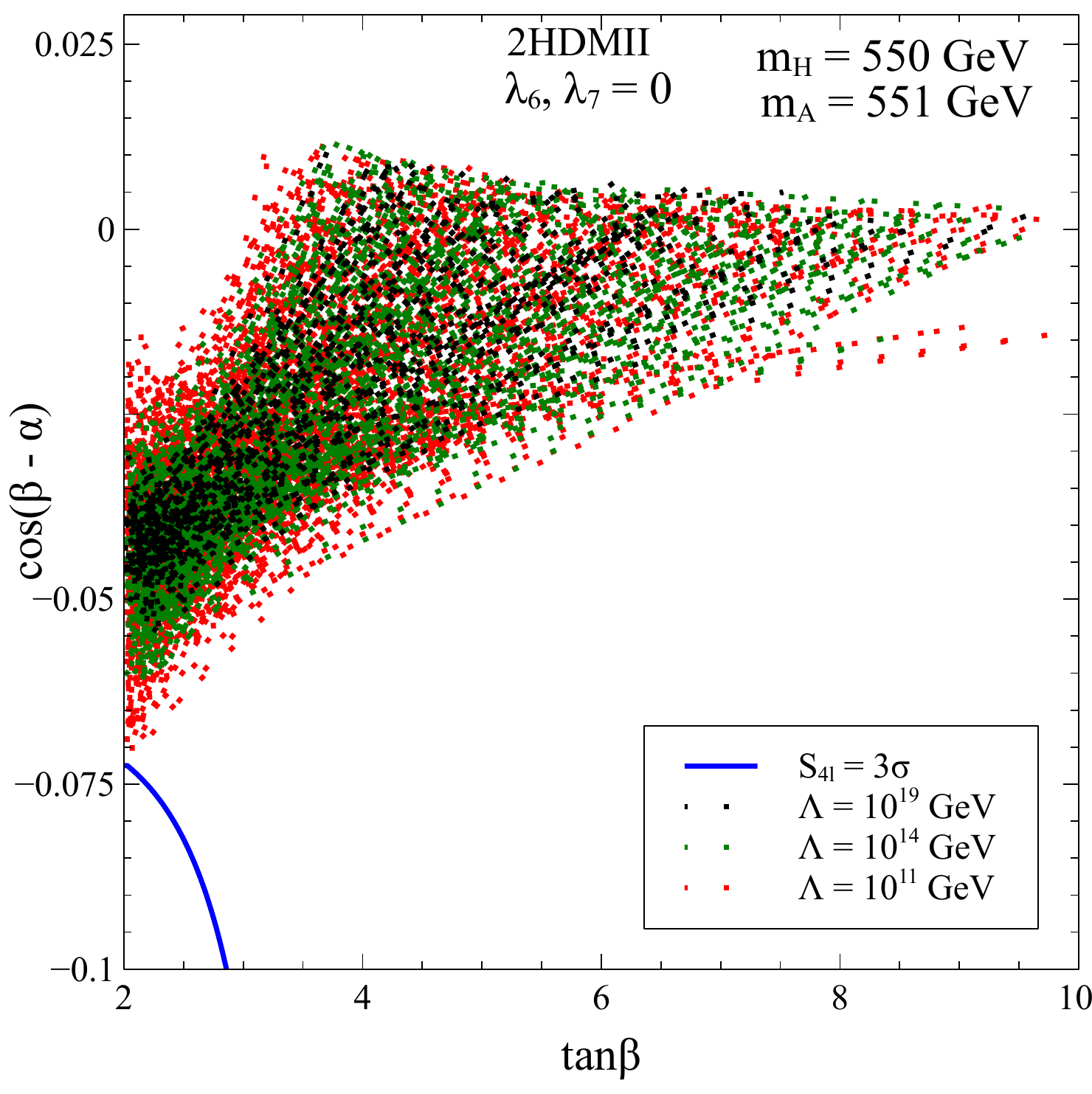}
\includegraphics[scale=0.48]{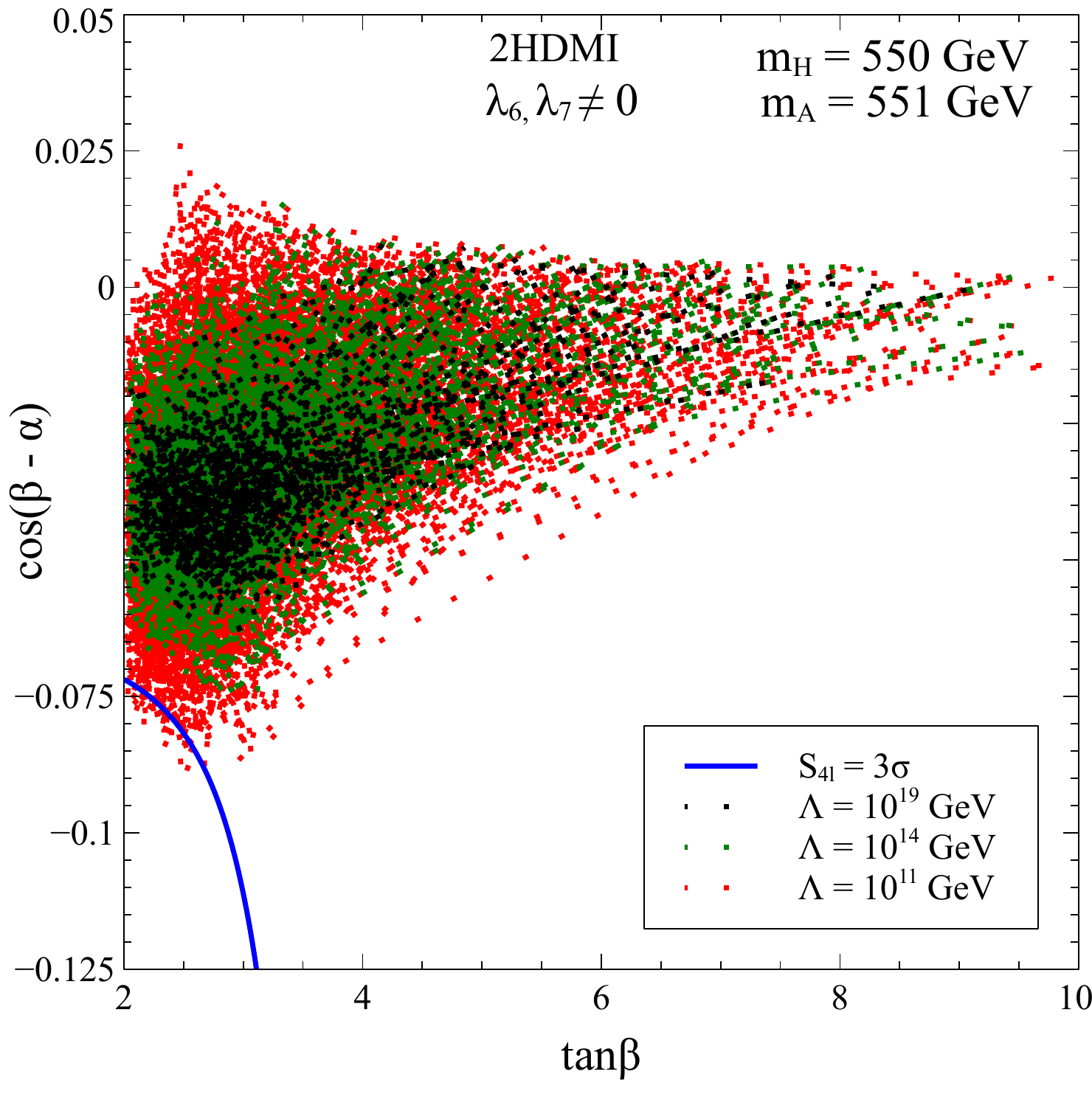}~~~ 
\includegraphics[scale=0.48]{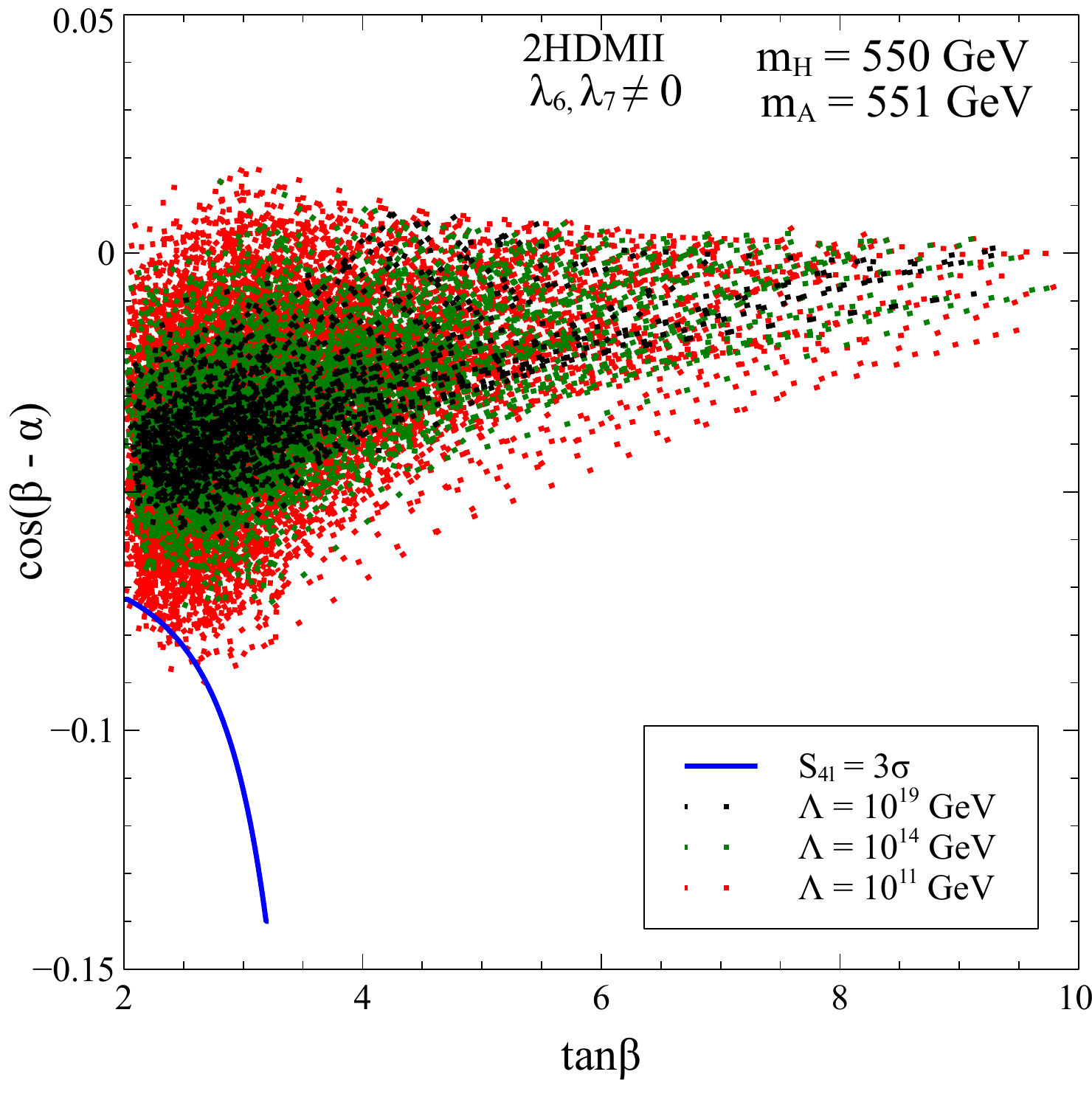}  
\caption{The paramater space in the tan$\beta$ vs. $c_{\b - \a}$ plane for $m_H = 550 \rm ~GeV$ and $m_A = 551$ GeV that allows for validity till $10^{11}$ GeV(red), $10^{14}$ GeV(green) and $10^{19}$ GeV(black). The region inside the blue curve corresponds to a signal significance greater than or equal to 3$\sigma$. The upper and lower plots are for $\l_6 = \l_7 = 0$ and $\l_6,\l_7 \neq 0$ respectively.}
\label{4l_550}
\end{center}
\end{figure}

Tables ~\ref{4l_TypeI} and ~\ref{4l_TypeII} contain the estimated CL for 
all the benchmarks. The following features thus emerge:\\
(i) The statistical significance diminishes as $m_H$ is increased. 
This is due to two reasons. First, the ggF cross section for a 
single $H$ drops. Secondly, the higher is $m_H$, the smaller is the upper limit
on $|\rm cos(\beta - \alpha)|$ consistent with high scale stability,
and hence, the lower is the $H \rightarrow Z Z$ branching ratio.\\
(ii) Type-I 2HDM offers a marginally higher significance as compared with Type-II. This is entirely attributed to the persistence 
of a slightly higher $H \rightarrow Z Z$ branching ratio in Type-I.\\
(iii) For $m_H \simeq 350$ GeV, an integrated luminosity of 100 $\rm fb^{-1}$ 
is sufficient to yield a 3$\sigma$ significance. \\
(iv) To observe an $H$ of mass around $\simeq$ 500 GeV that originates from
a 2HDM valid till $10^{19}$ GeV with a minimum of 3$\sigma$ confidence
level, one needs to gather 3000 $\rm fb^{-1}$
of data at the LHC. The statistical signifiance
decreases for higher masses. In short, the observability of a given $H$ can be improved by either 
lowering $m_H$ and holding the UV cutoff fixed or vice versa. This
interplay is illustrated in Fig.~\ref{4l_500} and Fig.~\ref{4l_550}.

Fig.~\ref{4l_500} corroborates the previous observation that an $H$ 
with $m_H = 500$ GeV can lead to a 3$\sigma$ signal at the LHC, consistently with perturbativity as well as a stable vaccum till $10^{19}$ GeV. This is true for both Type-I and Type-II 2HDM.
Note that the parameter space relaxes
upon the introduction of non-vanishing $\l_6$ and $\l_7$. This marginally helps 
in elevating the UV cutoff without compromising on the strength of the signal.
For $m_H = 550$ GeV, on the other hand, a 2HDM (of either Type-I or Type-II) cannot be be extrapolated beyond $10^{11}$ GeV if a 3$\sigma$ statistical significance has to be maintained. This is confirmed by an inspection of Fig.~\ref{4l_550}.

%\begin{figure} %[t]
%\begin{center}
%\includegraphics[scale=0.48]{m_4l_250.pdf}~~~ 
%\includegraphics[scale=0.48]{pTl1.pdf} 
%\caption{$m_{llbb}$ distribution for signal (green) and background (red) samples. The left and right
%panels correspond to BP1 and BP4 respectively.}
%\label{f:ST}
%\end{center}
%\end{figure}

We examine the prospects of reconstructing $A$ through the proposed
$l^+ l^- b \bar{b}$ final state in the following section.

\subsubsection{$p p \longrightarrow A \longrightarrow h Z \longrightarrow l^+ l^- b \bar{b}$}

In the absence of CP-violation (as assumed here), the $h Z$ pair production points towards a CP-odd
parent particle\cite{Khachatryan:2015lba}, and a peak in the invariant mass close to the afore mentioned $ZZ$ peak should be the somking gun signal of the near degeneracy of a scalar and a pseudoscalar. However, 
$p p \longrightarrow t\bar{t}$ generates the dominant background for this final state. Subleading contributions come from the production of $ZWW$ and $Z b \bar{b}$. Similar to the previous analysis, we adopt a $K$-factor = 1.5 for pseudoscalar production for all the benchmarks. The following cuts are applied during event-generation.

\textbf{Basic cuts:}
\begin{itemize}
\item $p_T^{l} \geq 10$ GeV, $p_T^{b} \geq 20$ GeV
\item $|\eta^{l}| \leq 2.5$, $|\eta^{b}| \leq 2.5$
\item $\Delta R_{ll} > 0.3$, $\Delta R_{lb} > 0.4$, $\Delta R_{bb} > 0.4$
\end{itemize}
On applying the above cuts, the NLO background cross section turns out to be $\simeq$ 32 pb. 
The following cuts are imposed for an efficient background rejection.

\textbf{Selection cuts:}
\begin{itemize}
\item C1: The invariant mass of the leptons satisfy $85.0 ~\rm GeV \leq m_{ll} \leq 100 ~\rm GeV$.
\item C2: The invariant mass of the b-jets satisfy $95.0 ~\rm GeV\leq m_{bb} \leq 155 ~\rm GeV$.
\item C3: The scalar sum of the transverse momenta of the leptons and b-jets satisfies $\sum_{l,b} p_T > (\sum_{l,b} p_T)_{\rm min}$.
\item C4: An upper bound on the missing transverse momenta, $\cancel{E_T} \leq 30$ GeV.
\item C5: The invariant mass of the $2l-2b$ system lies within the range $m_A - 30$ GeV $\leq m_{llbb} \leq m_A + 30$  GeV.
\item C6: $p_T$ of the reconstructed $Z$-boson satisfies $p_T^Z >  120$ GeV for BP5, and,  \\
$~~~~~~~~~~~ > 100$ GeV for the rest
\item C7: Upper bounds on the $p_T$ of the b-jets, $p_T^{b_1} > p_{T,\rm min}^{b_1}$.
\end{itemize}

The cuts on the $p_T$ of leading b-jet as well as on scalar the sum of the $p_T$ 
of the b-jets and leptons
are appropriately strengthened with increase in $m_A$.
We opt for $\{(\sum_{l,b} p_T)_{\rm min},p_T^{b_1}\}$ = $\{270 ~\rm GeV, 40 ~GeV\}$ for BP1, 
$\{320 ~\rm GeV, 40 ~GeV\}$ for BP2, 
$\{350 ~\rm GeV, 50 ~GeV\}$ for BP3 and BP4, and, 
$\{380 ~\rm GeV, 70 ~GeV\}$ for BP5.  
The selection cuts involve reconstructing the invariant masses of not only the decaying
$A$, but also of the $Z$ and the $h$, appropriately in each case. A lower limit on the scalar sum 
of the $p_T$ of the leptons and the $b$-hadrons also aids in incresing the significance.
All the $\cancel{E_T}$ in the signal is generated from mis-measurement of the momenta of the visible particles, thus generating a soft missing $\cancel{E_T}$ distribution. On the other hand, the corresponding background has a harder $p_T$ spectrum since the $t \bar{t}$ and $ZWW$ chennels always lead to neutrinos in the final state. Therefore, a suitable upper bound on the missing transverse energy reduces a portion of these backgrounds.

\begin{table}[h]
\centering
\begin{tabular}{|c c c c c c c c c|}
\hline
Benchmark  & $\sigma^{SC}_{S}$ (fb) & $\sigma^{SC}_{B}$ (fb) & $\mathcal{N}_S^{300}$ & $\mathcal{N}_B^{100}$ & $\mathcal{N}_S^{3000}$ & $\mathcal{N}_B^{3000}$ & $\rm CL_{100}$ & $\rm CL_{3000}$\\ \hline \hline
BP1a & 1.65 & 10.94 & 164.60 & 1094.05 & 4938.02 & 32821.48 & 4.64 & 25.41 \\
BP1b & 0.90 & 10.94 & 89.55 & 1094.05 & 2686.45 & 32821.48 & 2.60 & 14.26 \\ \hline
BP2a & 0.55 & 4.30 & 55.22 & 430.24 & 1656.63 & 12907.32 & 2.51 & 13.73 \\
BP2b & 0.28 & 4.30 & 27.92 & 430.24 & 837.64 & 12907.32 & 1.30 & 7.14 \\ \hline
BP3a  & 0.132 & 1.387 & 13.24 & 138.73 & 397.11 & 4161.95 & 1.07 & 5.88
\\
BP3b  & 0.076 & 1.387 & 7.63 & 138.73 & 228.91 & 4161.95 & 0.63 & 3.45
\\
BP3c  & 0.041 & 1.387 & 4.05 & 138.73 & 121.52 & 4161.95 & 0.34 & 1.86
\\ \hline
BP4a  & 0.066 & 0.632 & 6.56 & 63.22 & 196.86 & 1896.59 & 0.79 & 4.30\\
BP4b  & 0.044 & 0.632 & 4.35 & 63.22 & 130.50 & 1896.59 & 0.53 & 2.90\\
BP4c  & 0.031 & 0.632 & 3.08 & 63.22 & 92.53 & 1896.59 & 0.38 & 2.07\\ \hline
BP5a  & 0.021 & 0.334 & 2.07 & 33.37 & 62.19 & 1000.98 & 0.35 & 1.91\\
BP5b  & 0.010 & 0.334 & 1.05 & 33.37 & 31.38 & 1000.98 & 0.18 & 0.98\\
BP5c  & 0.005 & 0.334 & 0.54 & 33.37 & 16.27 & 1000.98 & 0.09 & 0.51\\ \hline
\end{tabular}
\caption{A record of the number of surviving events in the $A \rightarrow l^+ l^- b \bar{b}$ channel after the selection cuts at the $\sqrt{s} = 14 $ TeV LHC for a Type-I 2HDM. Here $\mathcal{N}_S^{100(3000)}$ and $\mathcal{N}_B^{100(3000)}$ and  respectively denote the number of events with $\mathcal{L} = 100(3000)$ $\rm fb^{-1}$. Besides,
$\rm CL_{100(3000)}$ denotes the confindence level for $\mathcal{L} = 100(3000)$ $\rm fb^{-1}$.}
\label{BP}
\end{table}

In this channel, too, Type-I fares slightly better than Type-II, much due to the same reason
outlined in preceding discussion. In this channel, The statistical significance of BP1-5
is also enhanced \emph{w.r.t} the $4l$ case, albeit marginally. 
The confidence level corresponding to $m_A$ = 500 GeV looms around 3$\sigma$, for both Type-I and Type-II.

\begin{table}[h]
\centering
\begin{tabular}{|c c c c c c c c c|}
\hline
Benchmark   & $\sigma^{SC}_{S}$ (fb) & $\sigma^{SC}_{B}$ (fb) & $\mathcal{N}_S^{300}$ & $\mathcal{N}_B^{100}$ & $\mathcal{N}_S^{3000}$ & $\mathcal{N}_B^{3000}$ & $\rm CL_{100}$ & $\rm CL_{3000}$\\ \hline \hline
BP3a  & 0.130 & 1.387 & 13.01 & 138.73 & 390.23 & 4161.95 & 1.06 & 5.78 \\
BP3b  & 0.075 & 1.387 & 7.49 & 138.73 & 224.80 & 4161.95 & 0.62 & 3.39\\
BP3c  & 0.040 & 1.387 & 3.98 & 138.73 & 119.30 & 4161.95 & 0.33 & 1.82\\ \hline
BP4a  &  0.065 & 0.632 & 6.45 & 63.22 & 193.61 & 1896.59 & 0.77 & 4.23 \\
BP4b  & 0.043 & 0.632 & 4.28 & 63.22 & 128.30 & 1896.59 & 0.52 & 2.85\\
BP4c  & 0.030 & 0.632 & 3.03 & 63.22 & 90.95 & 1896.59 & 0.37 & 2.04\\ \hline
BP5a  & 0.020 & 0.334 & 2.04 & 33.37 & 61.18 & 1000.98 & 0.34 & 1.88\\
BP5b  & 0.010 & 0.334 & 1.03 & 33.37 & 30.87 & 1000.98 & 0.18 & 0.96\\
BP5c  & 0.005 & 0.334 & 0.53 & 33.37 & 16.00 & 1000.98 & 0.09 & 0.50
\\ \hline
\end{tabular}
\caption{A record of the number of surviving events in the $A \rightarrow l^+ l^- b \bar{b}$ channel after the selection cuts at the $\sqrt{s} = 14 $ TeV LHC for a Type-II 2HDM. Here $\mathcal{N}_S^{100(3000)}$ and $\mathcal{N}_B^{100(3000)}$ and  respectively denote the number of events with $\mathcal{L} = 100(3000)$ $\rm fb^{-1}$. Besides,
$\rm CL_{100(3000)}$ denotes the confindence level at for $\mathcal{L} = 100(3000)$ $\rm fb^{-1}$.}
\label{BP}
\end{table}

\begin{figure} %[t]
\begin{center}
\includegraphics[scale=0.48]{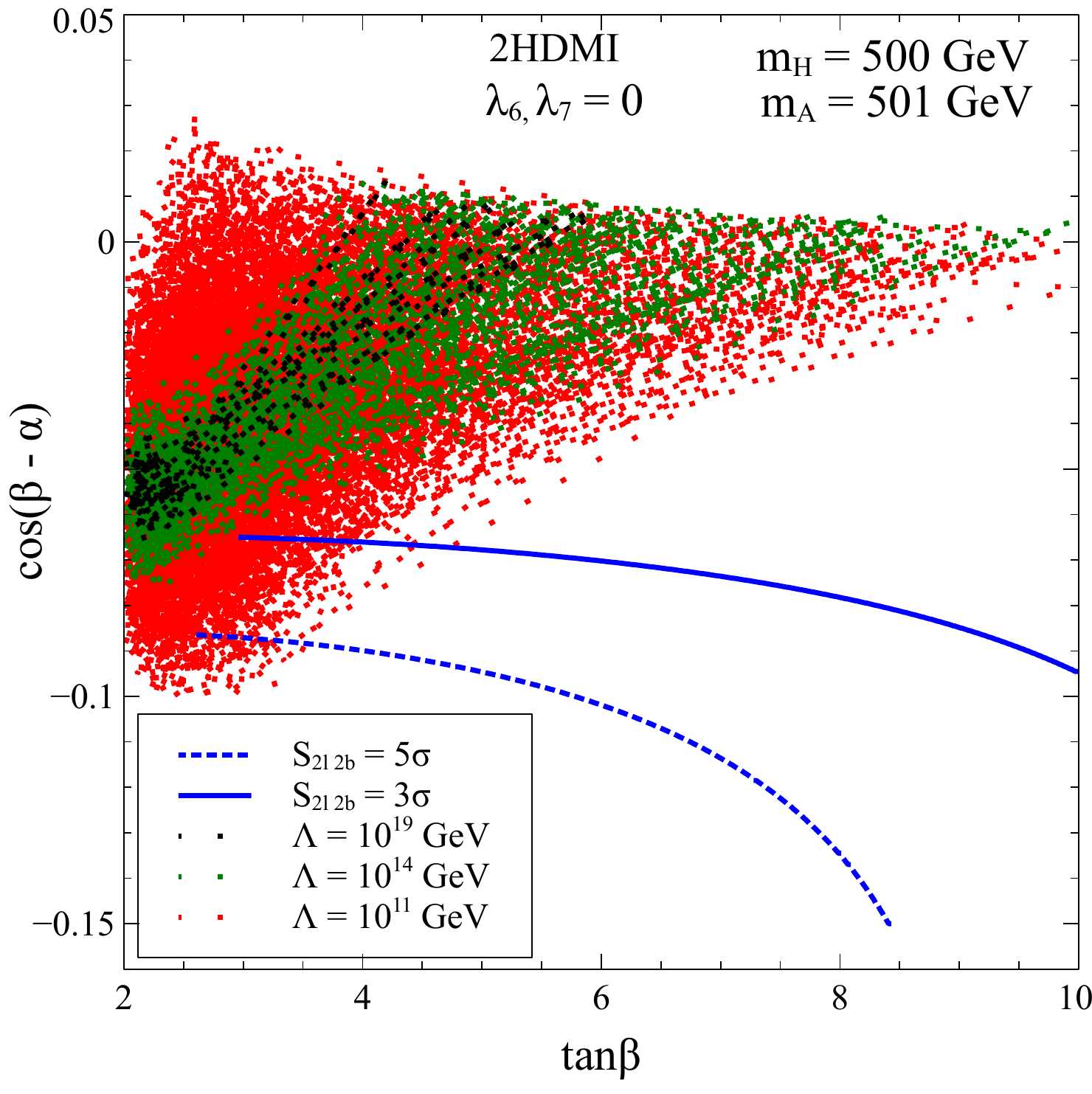}~~~ 
\includegraphics[scale=0.48]{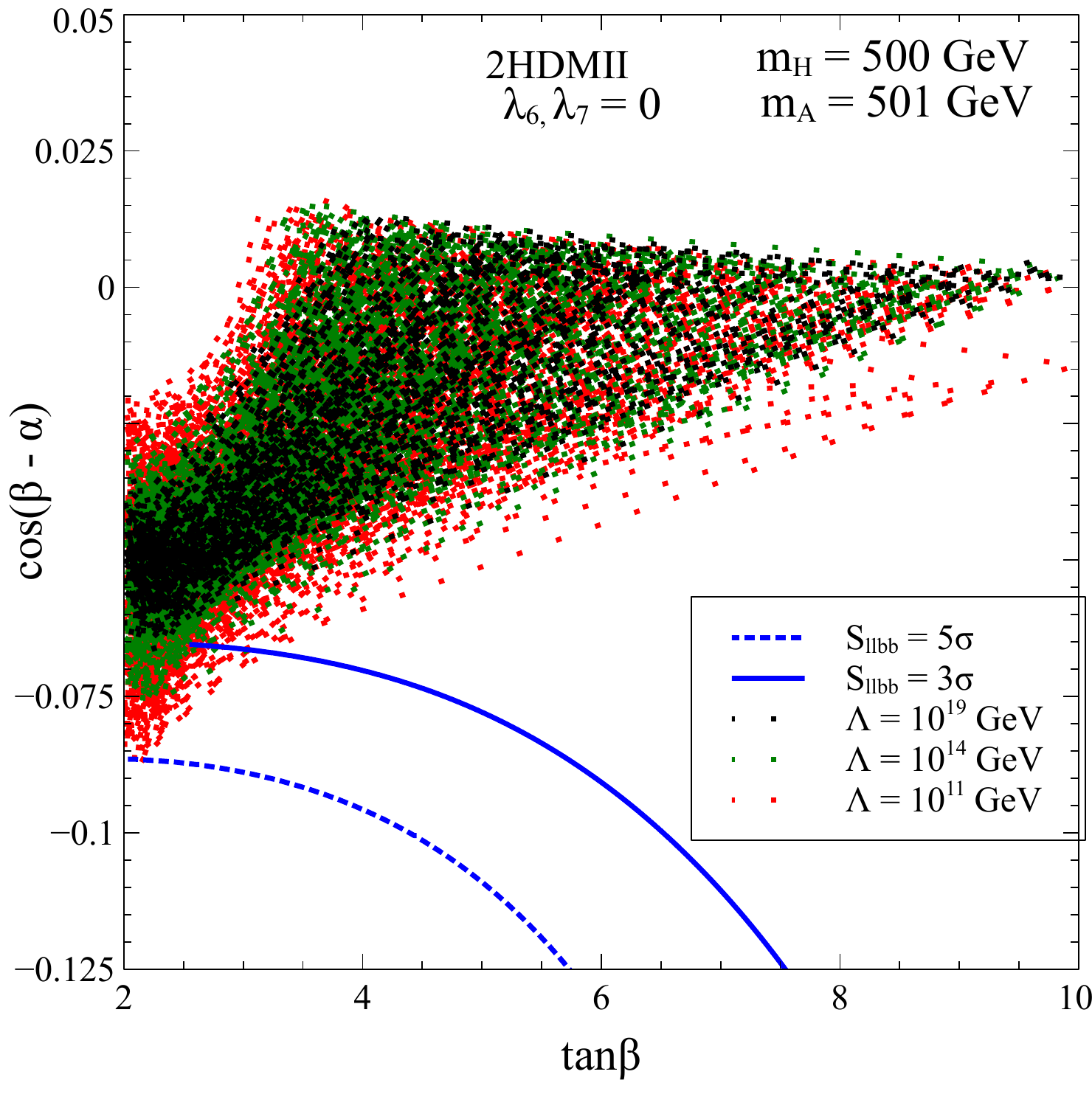}
\includegraphics[scale=0.48]{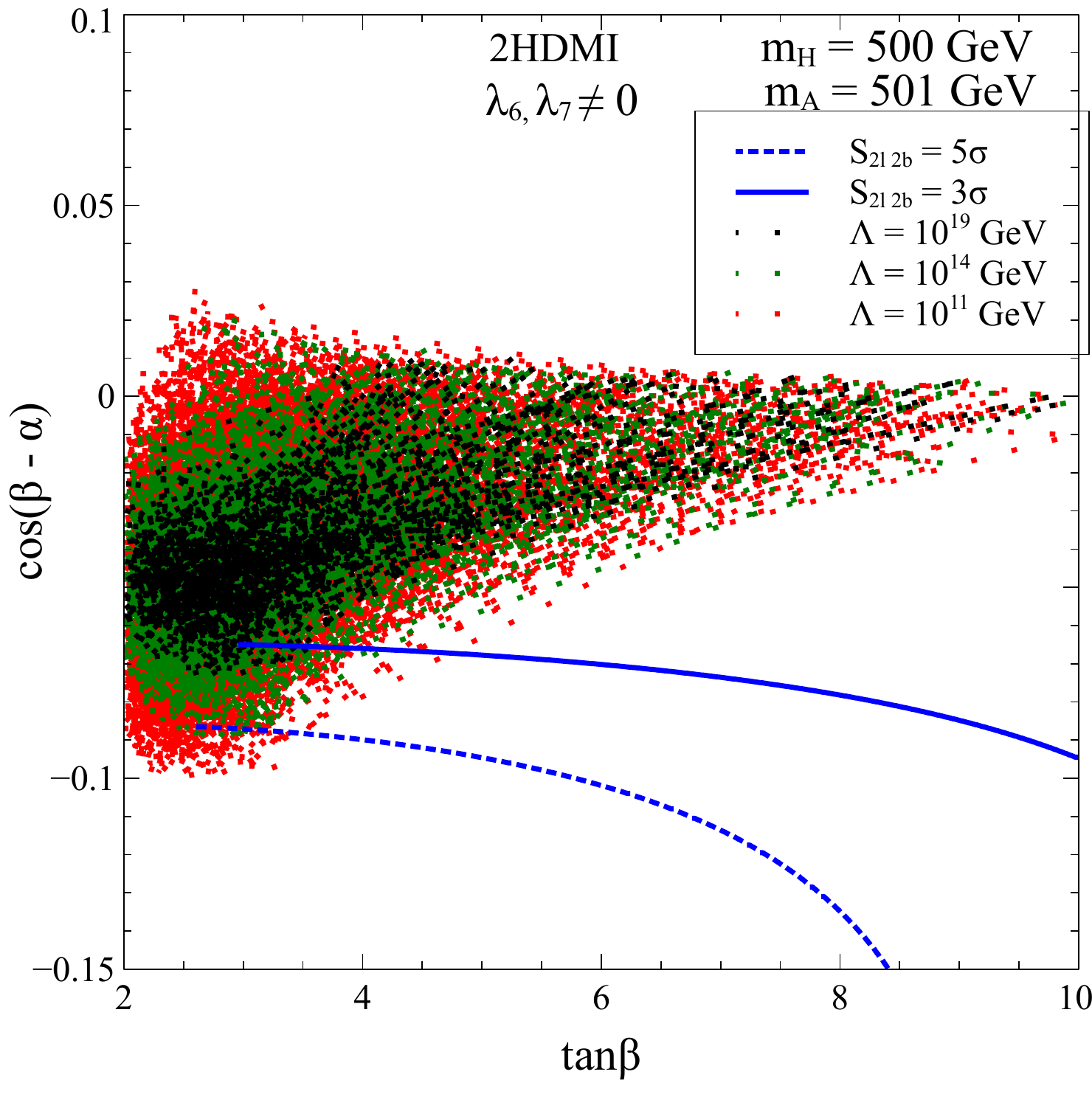}~~~ 
\includegraphics[scale=0.48]{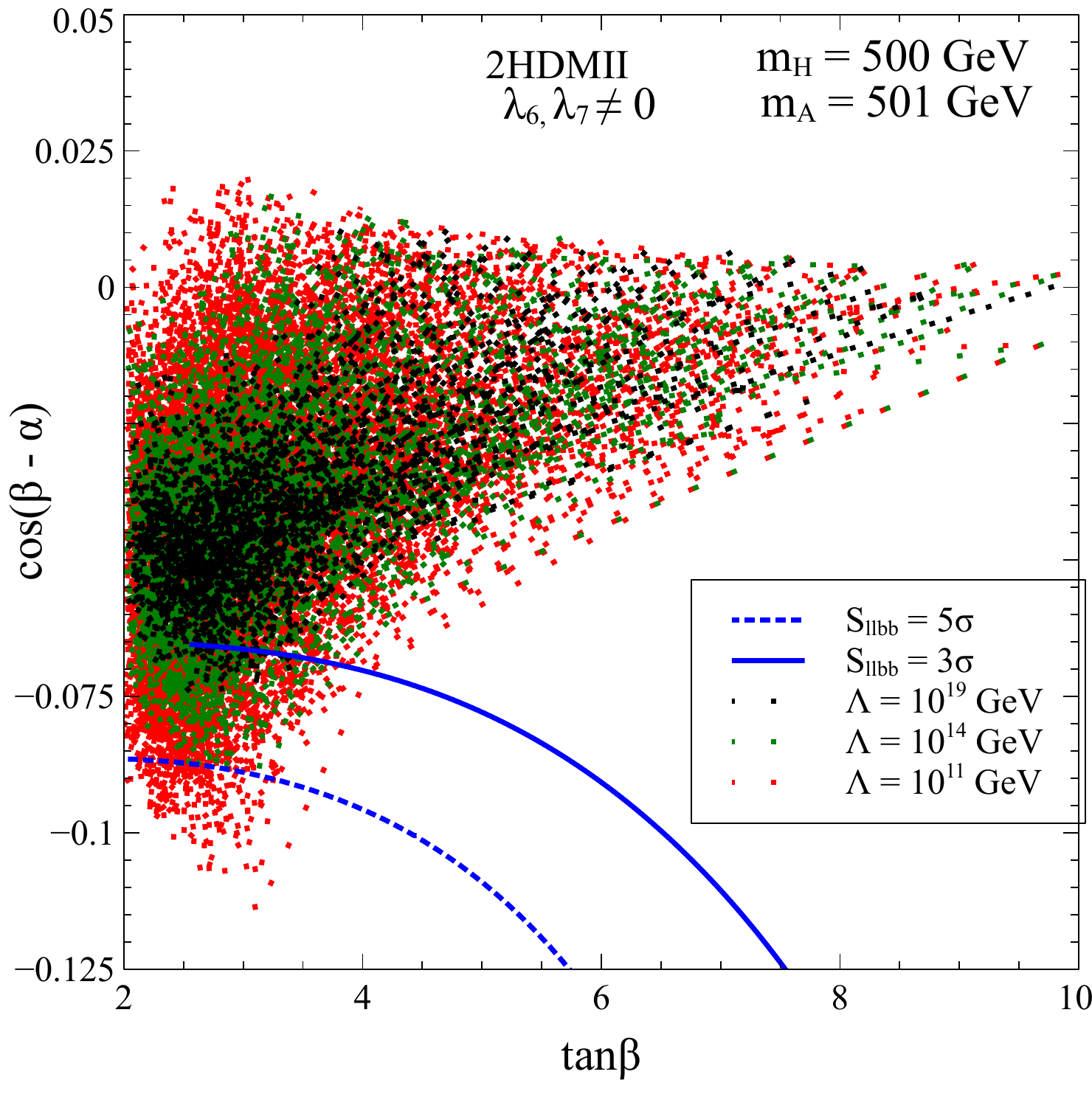}  
\caption{The paramater space in the tan$\beta$ vs. $c_{\b - \a}$ plane for $m_H = 500 ~\rm GeV$ and $m_A = 501$ GeV that allows for validity till $10^{11}$ GeV(red), $10^{14}$ GeV(green) and $10^{19}$ GeV(black). The region inside the solid (broken) blue curve corresponds to a signal significance greater than or equal to 3(5)$\sigma$. The upper and lower plots are for $\l_6 = \l_7 = 0$ and $\l_6,\l_7 \neq 0$ respectively.}
\label{llbb_500}
\end{center}
\end{figure}

\begin{figure} %[t]
\begin{center}
\includegraphics[scale=0.48]{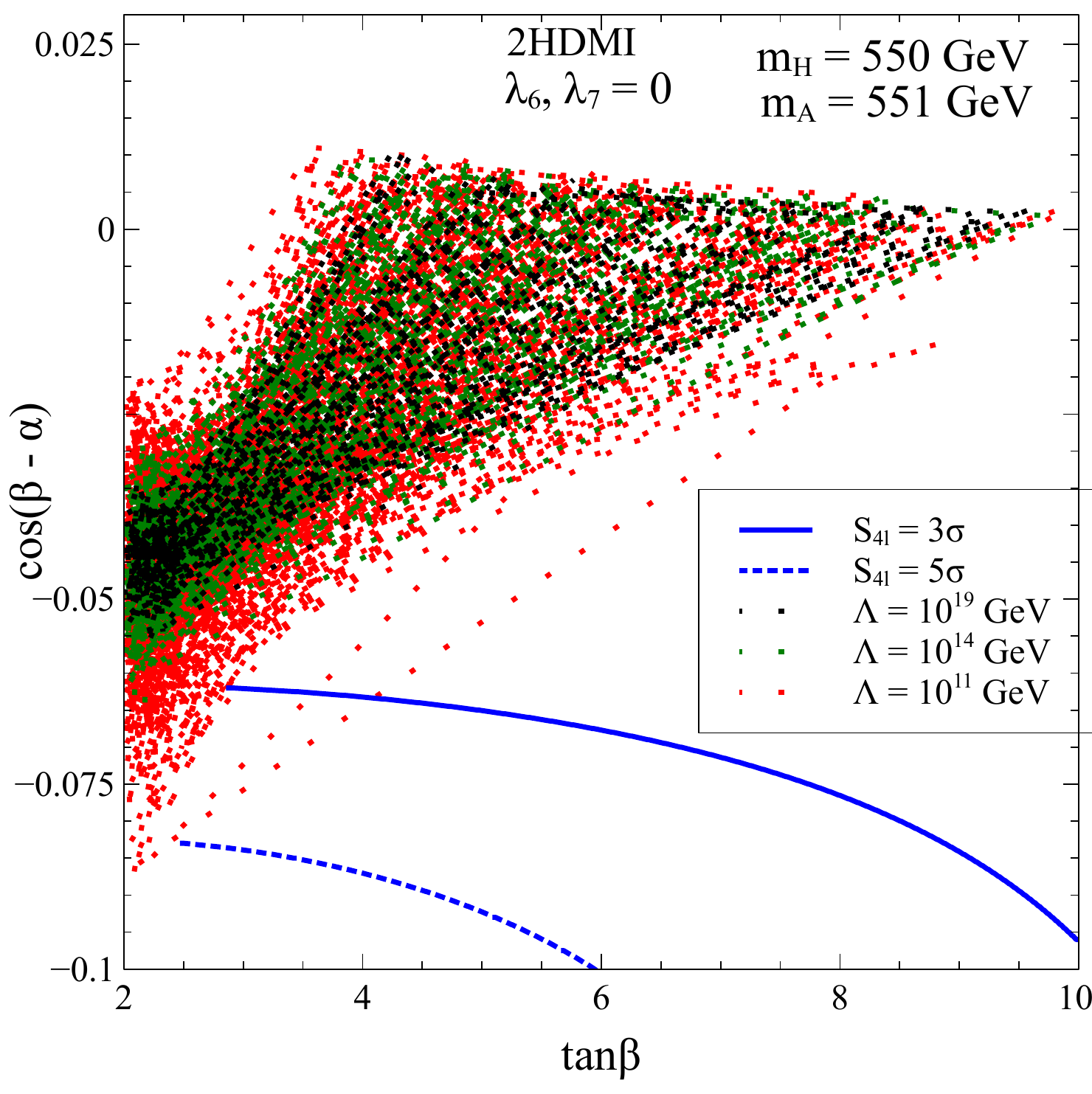}~~~
\includegraphics[scale=0.48]{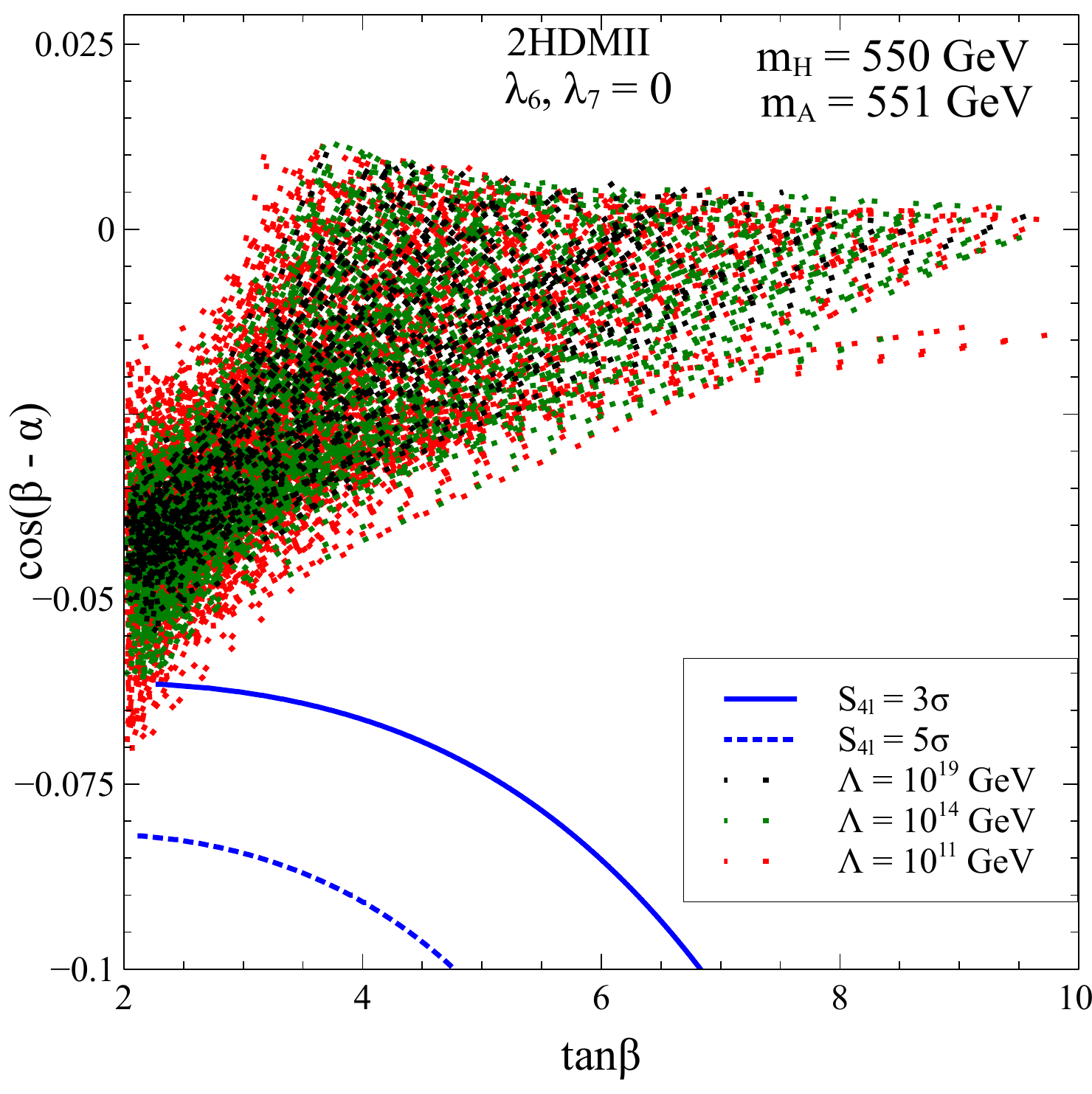}
\includegraphics[scale=0.48]{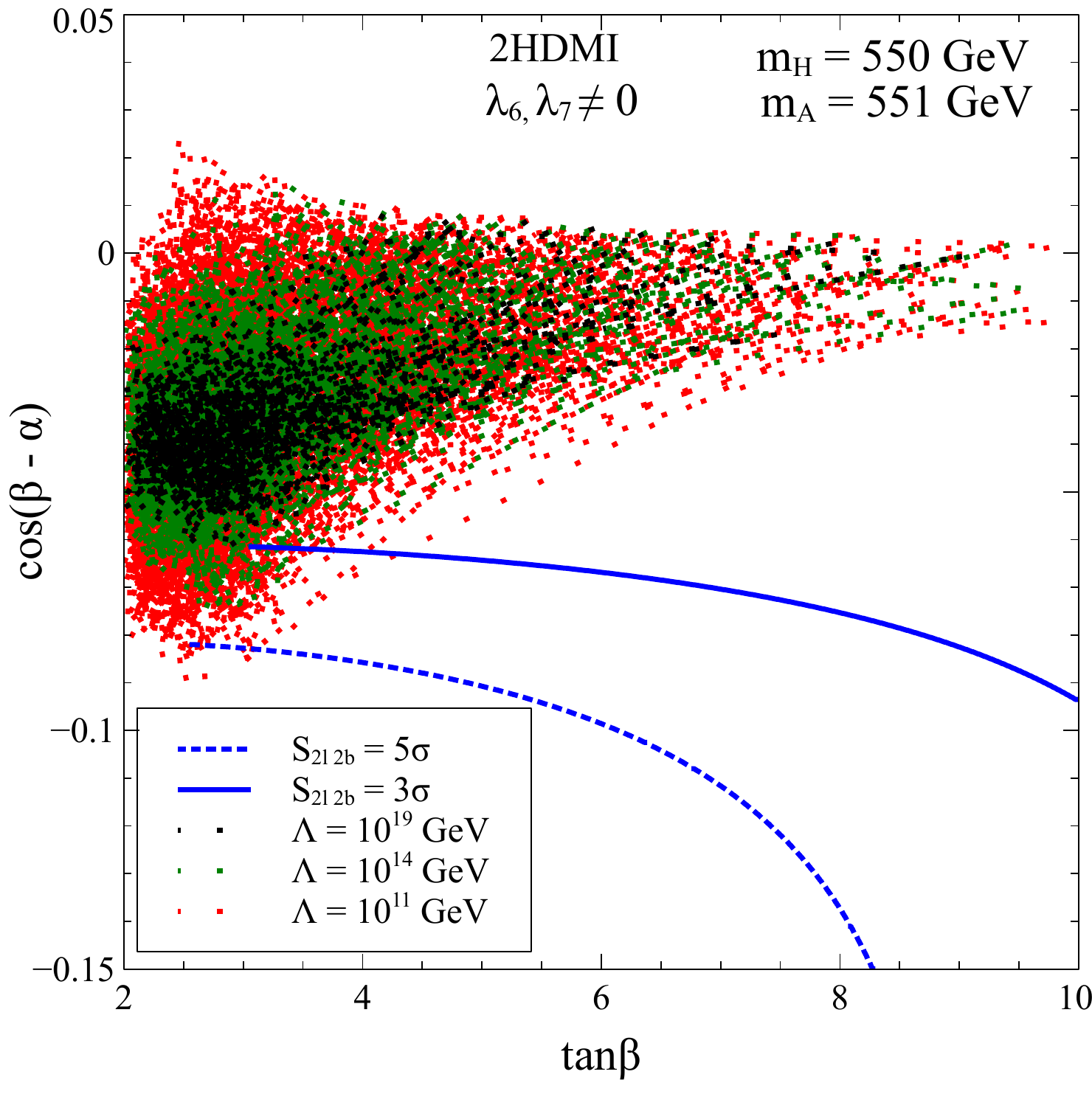}~~~ 
\includegraphics[scale=0.48]{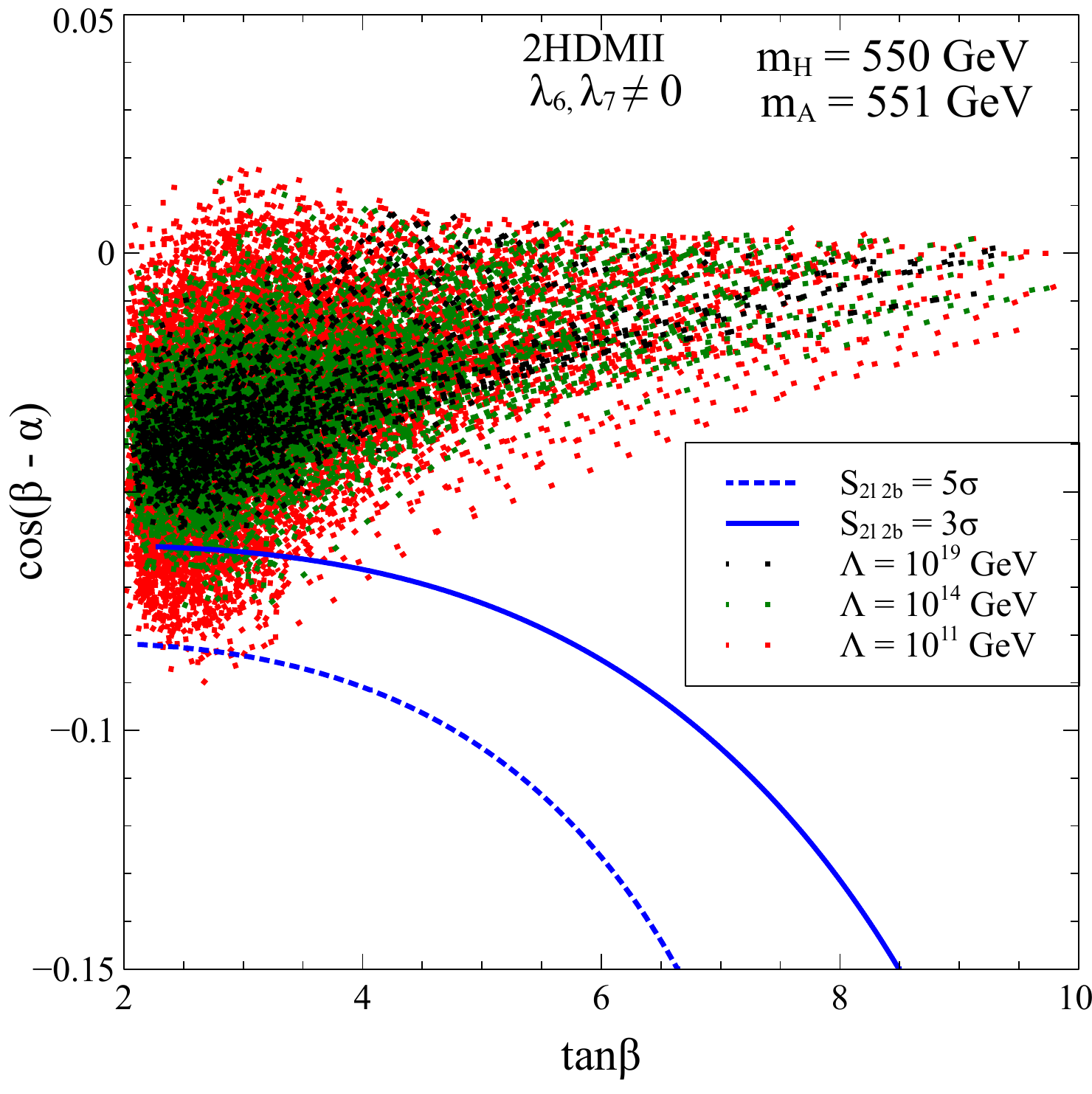}  
\caption{The paramater space in the tan$\beta$ vs. $c_{\b - \a}$ plane for $m_H = 550 ~\rm GeV$ and $m_A = 551$ GeV that allows for validity till $10^{11}$ GeV(red), $10^{14}$ GeV(green) and $10^{19}$ GeV(black). The region inside the solid (broken) blue curve corresponds to a signal significance greater than or equal to 3(5)$\sigma$. The upper and lower plots are for $\l_6 = \l_7 = 0$ and $\l_6,\l_7 \neq 0$ respectively.}
\label{llbb_550}
\end{center}
\end{figure}

A clearer picture regarding the observability of an $A$ of masses 500 GeV and 550 GeV emerge
upon inspection of Fig.~\ref{llbb_500} and Fig.~\ref{llbb_550} respectively. We display the 5$\sigma$ contour as well in case of the $l^+ l^- b \bar{b}$ channel. For $m_A = 550$ GeV with non-zero $\l_6$ and $\l_7$, the $l^+ l^- b \bar{b}$ channel offers sensitivity at the level of 3$\sigma$ for a scenario valid till 10$^{14}$ GeV or even higher. On the contrary, the corresponding cut-off cannot be pushed above $10^{11}$ GeV if one demands
similar observability in case of the 4$l$ final state from $H$-decay. Overall, a violation of the $Z_2$ symmetry via $\l_6$ and $\l_7$
aids to the effort of observing a 2HDM valid up to high cut-off scales.

It is mentioned that the analysis for this channel is subject to uncertainties, albeit small, that are introduced while estimating the background cross section. Upon considering the errors in the 
$t \bar{t}$ production rates and the background NLO K-factors\cite{Alwall:2014hca}, the total background cross section can deviate up to $\simeq \pm 20\%$. This, however, does not modify the overall conclusions made in this section.

\section{Prospects at other colliders}\label{other}

With the prospects of observing non-standard scalars with masses above the 500 GeV range at the LHC turning bleak, we resort to future lepton colliders for better observability. These include not only the $e^+ e^-$ colliders, but also a muon collider\cite{PhysRevD.88.115003}. 

The principal heavy Higgs production channels at the $e^+ e^-$ machine are those of associated production (VH) and Vector-Boson-Fusion (VBF)\cite{Hodgkinson:2009uj}. The production rate in both of these modes is controlled by the value of $\rm cos(\b-\a)$. As elaborated in the previous sections, $\rm cos(\b-\a)$ is tightly
bounded by the requirement of a stable vacuum till the Planck scale. In addition, the maximum $\sqrt{s}$ proposed for the ILC is 1 TeV\cite{Behnke:2013lya} which hampers a probe of heavy scalars due to
kinematical limitations. For instance, the VH production cross section for an $H$ of mass 600 GeV could be at most $\simeq$ 0.01 fb in a ILC with $\sqrt{s} = 1$ TeV. This does not result in the requisite signal significance when the backgrounds are estimated and the cut efficiencies are folded in.

\subsection{$\mu^+ \mu^-$ collisions and radiative return}\label{radret}
A particularly interesting process in a muon collider is one of
radiative return (RR)\cite{PhysRevD.91.015008}, where one does not need to know the mass of the resonantly produced scalar precisely. In our context, the processes under consideration are
\be
\mu^+ \mu^- \longrightarrow H ~\gamma, A ~\gamma 
\ee
Note here that $H/A$ can be produced in association with a $\gamma$
in t-channel $\mu^+ \mu^-$ annihilations.
When the center of mass energy of the muon
collider is above the heavy resonance, the photon emission from the initial state provides an opportunity to reconstruct the mass of the heavy scalar or pseudoscalar. For this, one need not know the mass of the (unknown) heavy resonance. The final state then consists of a soft photon and other visible products exhibiting an invariant mass peak. The closer is the mass of the heavy scalar to the centre-of-mass (COM) energy of $\mu^+ \mu^-$ collisions, the higher the cross section. 

Thus tagging a heavy scalar state from invariant mass peak of its decay product can help us in reducing the background
and increasing the statistical significance. Moreover, in order to obtain information on the CP of the heavy resonance, the CP-even and the CP-odd states must be allowed to decay in different final states following their production through RR. We can propose $H \longrightarrow Z Z \longrightarrow 4 l$ and $A \longrightarrow h Z \longrightarrow l^+ l^- b \bar{b}$, which resemble the signals studied in the previous sections, and distinguish the cP-even scalar from the CP-odd one. In order to study the observability of the benchmarks BP3a - BP5c in RR, we choose the COM energy of the $\mu^+ \mu^-$ collisions to be just 10 GeV above $m_H$, in each case. For BP3a, the RR cross section for $H$ production is $\simeq$ 1.3 fb for a Type-II 2HDM. Upon multiplying by the branching ratios corresponding to $H \rightarrow Z Z$ and $Z \rightarrow l l$, the corresponding cross section for the $4l + \gamma$  final state turns out to be $\mathcal{O}$(${10^{-4}}$) (fb). The cross section for the $l^+ l^- b \bar{b} + \g$ final state could still be $\mathcal{O}$(${10^{-3}}$) (fb). However, it will ultimately get reduced when kinematical cuts are applied. In a Type-II 2HDM,
though the $\mu\mu H$ coupling is proportional to tan$\beta$, 
opting for a higher value of tan$\beta$ does not help in this regard, since in that case, the allowed value of $|\rm cos(\b-\a)|$ decreases owing to the demand of validity till high scales (see Fig.). This diminishes the $H \rightarrow Z Z$ and $A \rightarrow h Z$ branching ratios, and ultimately, leads
to further lower rates. The other BPs too predict negligibly small RR rates for both Type-I and Type-II.  With such meagre RR rates in $4l + \gamma$ as well as $l^+ l^- b \bar{b} + \g$ channels, chances of observing the heavy resonances are obliterated. 

Still promising could the fermionic decay channels of $H/A$ in this regard. For instance, the $b \bar{b} A$ coupling in a Type-(I)II 2HDM is proportional to cot$\beta$(tan$\beta$) and for sufficiently small $|\rm cos(\b-\a)|$, the fermionic couplings of $H$ and $A$ are nearly equal.
The advantage of a muon collider over the LHC is that the $b \bar{b}$ final state can rise above the background more effectively. As we shall see below, this enhances the mass reach.
 
One can thus probe the observability of the heavy scalars in the $\mu^+ \mu^- \rightarrow H/A \g \rightarrow b \bar{b} \g$\footnote{In view of the high t-Yukawa coupling, one could also look at $\mu^+ \mu^- \rightarrow H/A \g \rightarrow t \bar{t} \g$ in principle. However that channel will ultimately lead to lesser rates compared to the $b \bar{b}$ mode owing to the smaller $t \bar{t}$ branching fraction.}. It is readily seen that for tan$\beta > 1$,
Type-II has higher production rates of $H/A$ through RR compared to Type-I. This could give a handle in distinguishing between Types I and II. Therefore, to test the potency of RR in the $H/A \rightarrow b \bar{b}$ mode, we tabulate two additional benchmarks, as shown in Table ~\ref{BP_radret}.

\begin{table}[h]
\centering
\begin{tabular}{|c c c c c|}
\hline
Benchmark  & $\sqrt{s}$ (GeV) & tan$\beta$ & $m_{H}$(GeV) & $m_{A}$(GeV) \\ \hline \hline

BP6 & 500 & 12 & 492 & 493 \\ \hline
BP7 & 1000 & 12 & 992 & 993 \\ \hline

\end{tabular}
\caption{The values of $m_H$, $m_A$ and tan$\beta$ chosen to probe the radiative return channel.The values of $\sqrt{s}$ are also shown.
}
\label{BP_radret}
\end{table}

The values of the other 2HDM parameters have been fixed appropriately so as to ensure
stability till the Planck scale. For instance, we chose $m_{12} = 150$, $c_{\b - \a} = 0.01$ and  $m_{12} = 500$, $c_{\b - \a} = 0.005$ for BP6 and BP7 respectively. We take 500 GeV and 1 TeV to be COM energy for these two cases. 
Accordingly, $\sqrt{s} - m_{H/A}$ is maintained around $\sim$ 7 GeV to maximize the efficiency of the radiative return mechanism. In addition, we have purposefully chosen a somewhat large value
for tan$\beta$ to elevate the $H/A$ production rate to the order of 10 fb. Moreover, we also get a sizeable branching ratio for the $H/A \rightarrow b \bar{b}$ channel for both BP6 and BP7 ($> 70 \%$). 

The SM background comes from the processes $\mu^+ \mu^- \rightarrow b \bar{b}$ and $\mu^+ \mu^- \rightarrow b \bar{b} \g$. 
The cut, $m_H - 30 ~\text{GeV} < m_{bb} < m_H + 30 ~\text{GeV}$ on the invariant mass of the $b$-pair is imposed. The softness of the photon in the case of RR can be exploited to reduce the background by putting an upper bound on the photon $p_T$, which we take to be 30 GeV.
Effects arising out of smearing the photon-energy
are small, so we keep the photon-energy same as the simulated value.
The confidence levels obtained for BP6 and BP7 are listed in Table~\ref{sig_radret}. 
 
\begin{table}[h]
\centering
\begin{tabular}{|c c c c c c c c c|}
\hline
Benchmark   & $\sigma^{SC}_{S}$ (fb) & $\sigma^{SC}_{B}$ (fb) & $\mathcal{N}_S^{500}$ & $\mathcal{N}_B^{500}$ & $\mathcal{N}_S^{1000}$ & $\mathcal{N}_B^{1000}$ & $\rm CL_{500}$ & $\rm CL_{1000}$\\ \hline \hline
BP6  & 2.02 & 32.22 & 1011.28 & 16110.05 & 2022.56 & 32220.08 & 7.72 & 10.92\\ \hline
BP7  & 0.26 & 2.52 & 133.92 & 1264.28 & 267.85 & 2528.57 & 3.58 & 5.06\\ \hline
\end{tabular}
\caption{Number of signal and background surviving events in the radiative return process at the muon collider. Here $\mathcal{N}_S^{500(1000)}$ and $\mathcal{N}_B^{500(1000)}$ and  respectively denote the number of events $\mathcal{L} = 500(1000)$ $\rm fb^{-1}$. Besides,
$\rm CL_{500(1000)}$ denotes the confindence level at $\mathcal{L} = 500(1000)$ $\rm fb^{-1}$.}
\label{sig_radret}
\end{table}

Table~\ref{sig_radret} shows that it is possible to experimentally observe an $H$ as heavy
as 1 TeV through radiative return. The corresponding signal rates are almost identical for a
near degenerate $A$ decaying to $b \bar{b}$, and thus, are not separately shown. Thus, 
radiative return in the $b \bar{b}$ channel does succeed in predicting abundant signal events 
in case of heavy scalars. This is reflected by a sizable statistical significance of $\sim$ 5$\sigma$ that can be obtained in case of a scalar of mass 1 TeV when the $\mu^+ \mu^-$ machine 
is operated at an integrated luminosity of 1000 fb$^{-1}$. More importantly, this is found to be in perfect agreement with high-scale stability and perturbativity up to $M_{Pl}$.
However in this channel, one faces the difficulty in distinguishing between a $b \bar{b}$ resonance that comes from an $H$ and one coming from $A$. This is in sharp contrast with the results 
obtained in case of the 14 TeV LHC. Over there, though the $CP$ of the scalar can be tagged, 
its observability does not exceed 3$\sigma$ in terms of confidence level for masses beyond 500 GeV.

%Table~\ref{sig_radret} reports that it is possible to experimentally observe an $H$ as heavy
%as 1 TeV through radiative return. The corresponding signal rates are almost identical for a
%near degenerate $A$ decaying to $b \bar{b}$, and thus, are not separately shown. Thus, while radiative return in the $b \bar{b}$ channel succeeds in predicting abundant signal events, one faces the difficulty in distinguishing between a $b \bar{b}$ resonance that comes from an $H$ and one coming from $A$. This coincidence of two invariant mass distributions can be avoided if the CP-even and the CP-odd states are allowed to decay in different final states following their production through radiative return. Similar to the analysis for the LHC, the final states of $4l$ and $2l 2b$ in association with a photon with each case can be looked at to tag the $H$ and $A$ respectively, in the context of $\mu^+ \mu^-$ collisions also. Any such effort would call for
%maximizing the prefactor tan$\beta$ $c_{\b - \a}$ within the parameter space catering to high scale validity. 

\section{Summary and conclusions.}\label{con} 

By virtue of the additional bosonic fields, a 2HDM ensures the stability of the EW vacuum
till a cut-off scale all the way up to the Planck scale. This holds true even after switching between
the Type-I and Type-II cases. However stringent constraints apply on the parameter space in the
process. This is especially true when vacuum stability and perturbative unitarity are demanded up to the Planck scale. Then, the couplings of the non-standard scalars to other bosonic states become very
small because of suppressed cos($\b - \a$). In addition, the mass spectrum of the non-standard scalar bosons becomes quasi-degenerate. These constraints limit the observability of such a 2HDM 
at colliders.

We have studied in detail the interplay between high-scale validity and the discernability of the scenario at the LHC and at a future muon collider. In the LHC,
signatures of the the CP-even boson $H$ and CP-odd boson $A$ are studied through their decays
into the $4l$ and $l^+ l^- b \bar{b}$ channel respectively. The search turns challenging 
due to the stringent upper bound on cos($\b - \a$). A sizable signal significance demands an upper bound on tan$\beta$, contrary to high scale validity constraints, where no such bound is predicted. An analysis at the 14 TeV LHC including detector effects
reveals that $H$ and $A$ of masses around 500 GeV can be simultaneously observed in their respective channels with at least 3$\sigma$ confidence when the integrated luminosity is 3000 fb$^{-1}$. The observability improves upon de-escalating the cut-off scale, attaining 5$\sigma$ 
statistical significance becomes possible when the cut-off is near $10^{11}$ GeV. 

Radiative return at the muon collider yields sizable production rates of $H$ or $A$. 
We have studied the observation their prospects through their subsequent decay to the 
$b \bar{b}$ final state. Contrary to the results obtained for the LHC, the 
$\mu^+ \mu^-$ machine can lead to a 5$\sigma$ statistical significance even if the scalar mass
is 1 TeV. Thus a certain complimentarity of roles between the LHC and a muon collider is noticed.
The former has relatively lower mass reach but clearly differentiates the $H$-peak from the $A$-peak, while the latter loses this distinction by being forced to look at the $b \bar{b}$ decay mode, though up to higher (pseudo)scalar masses.

\section{Acknowledgements}

We thank Subhadeep Mondal and Jyotiranjan Beuria for useful discussions. This work was partially supported by funding available from the Department of Atomic
Energy, Government of India, for the Regional Centre for Accelerator-based Particle Physics (RECAPP), Harish-Chandra Research Institute.

%%%%%%%%%%%%%%%%%   References %%%%%%%%%%%%%%%%%%%%%%%%%%%%%%%%%%%%
\bibliographystyle{JHEP}
\bibliography{ref.bib}        
\end{document}